\documentclass[prb,eqsecnum,showpacs,preprint]{revtex4}
\usepackage{amssymb,amsfonts,amsmath,bm,dcolumn,latexsym,graphicx,axodraw}
\font\bba=msbm10 scaled 1200
\font\bbb=msbm8 
\font\bbc=msbm6 
\newfam\bbfam
\textfont\bbfam=\bba
\scriptfont\bbfam=\bbb
\scriptscriptfont\bbfam=\bbc
\def\bb{\fam\bbfam\bba}
\def\R{{\bb R}}
\def\Z{{\bb Z}}

\begin{document}
\title{Non-Perturbative Renormalization Group for Simple Fluids}
\date{\today}
\author{Jean-Michel Caillol}
\affiliation{Laboratoire de Physique Th\'eorique \\
UMR 8627, B\^at. 210 \\
Universit\'e de Paris-Sud \\
91405 Orsay Cedex, France}
\email{Jean-Michel.Caillol@th.u-psud.fr}
\begin{abstract}
We present a new non perturbative renormalization group for classical simple fluids. The theory is built  in the Grand Canonical ensemble and in the framework of two equivalent scalar field theories as well. The exact mapping between the three renormalization flows is established rigorously. In the Grand Canonical ensemble the theory may be seen as an extension of the Hierarchical Reference Theory (L. Reatto and A. Parola, \textit{Adv. Phys.}, \textbf{44}, 211 (1995)) but however does not suffer from its shortcomings at subcritical temperatures. In the framework of a new canonical field theory of liquid state developed in that aim our construction identifies with the effective average action approach developed recently (J. Berges, N. Tetradis, and C. Wetterich, \textit{Phys. Rep.}, \textbf{363} (2002)).\\\\
\textbf{keywords} : Critical Phenomena, Renormalization Group, Theory of Liquids, Hierarchical Reference Theory, Liquid-Vapor Transition, Kac Model.
\end{abstract}
\maketitle
\section{\label{intro}Introduction}
During the last twenty years the Wilson approach\cite{Wilson} to the renormalization group (RG) has been the subject of a revival in both statistical physics and field theory. Two main formulations of the non perturbative renormalization group (NPRG) have been developed in parallel. In the first one, a continuous realization of the RG transformation of the Hamiltonian $\mathcal{H}_k[\varphi]$ is made and no expansion is involved with respect to some small parameter of the action $\mathcal{H}_k[\varphi]$. At scale-$k$ (in momentum space) the high energy modes $\widetilde{\varphi}_q$, $q>k$, have been integrated out in
$\mathcal{H}_k[\varphi]$  which requires the implementation of some cut-off of the propagator aiming at separating slow ($q<k$) and fast ($q>k$) modes. The flow of the action is governed by either by the Wilson-Polchinski equation\cite{Wilson,Wegner,Polchinski} in case of a smooth cut-off or the  Wegner-Houghton\cite{Wegner2} equation in case of a sharp cut-off. These equations, due to their complexity, call for the use of approximation and/or truncation methods. We refer to the recent review of Bagnuls and Bervillier\cite{Bervillier} of a detailed discussion of this version of the NPRG. The second formulation, called the effective average action approach, was developed after the seminal works of Nicoll and Chang (sharp cut-off version)\cite{Nicoll} and Wetterich (smooth cut-off version)\cite{Wetterich0}. This method implements on the effective action $\Gamma_k[\Phi]$ -the Gibbs free energy of the fast modes $\widetilde{\Phi}_q$ , $q>k$- rather than on   $\mathcal{H}_k$ the ideas of integration of high-energy modes that underlies any RG approach. The flow of $\Gamma_k$ results in  equations which can be solved under the same kind of approximations than those used for the Wilson-Polchinski or Wegner-Houghton equations. Recent reviews and lectures on this second approach are  available\cite{Wetterich,Delamotte} and will be consulted with profit. It must be stressed that these two approaches of the NPRG are of course equivalent as long as the exact equations are considered but may yield different results when approximations are introduced at some stage\cite{Morris,Ellwanger}. The effective average action method has been applied successfully to compute with accuracy the critical exponents of the Ising model non perturbatively  for instance\cite{Wetterich,Delamotte2,Delamotte3} and also to tackle with other highly non-trivial problems of condensed matter physics such that reaction-diffusion processes\cite{Delamotte4}, frustrated magnets\cite{Tissier} or the Kosterlitz-Thouless transition\cite{Gersdorff}, etc.

In this paper we implement the method of the effective average action for simple classical liquids. This can be done either in the framework of a  statistical field theory of liquids\cite{Brilliantov,Song,Cai-Mol,Cai-Pat1,Cai-Pat2} or in the usual Grand Canonical (GC) formalism\cite{Hansen,Stell1,Stell2} of classical statistical mechanics. The first RG approach to the theory of simple fluids and mixtures was  proposed more than twenty years ago in a series of outstanding papers by Reatto, Parola and co-workers who developed the so-called Hierarchical Reference Theory (HRT) which gives an accurate description of the thermodynamics and the structure of a wide class of fluids close to and away from the critical region (for a review see reference\cite{HRT}). It turns out\cite{HRT} that HRT is equivalent to the NPRG with sharp cut-off of Nicoll and Chang \cite{Nicoll}.
The HRT will be obtained here as a highly singular limit of a regular NPRG equation, i.e. as an ultra-sharp cut-off limit. We claim that the theory discussed in this paper is free of the flaws which affect HRT in the sub-critical region, i.e. that the coexistence and spinodal curves merge into a single curve\cite{HRT2,Reiner1,Reiner2}.  It is likely -and some arguments will be developed in that sense- that the singularity of the limit in which HRT is obtained is the cause of its shortcomings giving little hope to improve the theory by considering more sophisticated approximations and/or truncations such that those proposed recently\cite{Reiner3}; see however a little known paper of Parola\cite{Parola}.

Our article is organized as follows. In section\ (\ref{RG-GC}) we present our NPRG theory in the framework of the Grand Canonical of classical statistical mechanics. We consider only the case of a fluid of hard spheres with additional attractive pair interactions although the generalization to multicomponent fluids would be trivial. After some prolegomena aimed at resuming the basics notions of the theory of liquids and at fixing our notations (section\ (\ref{Prolego})) the exact RG flow of the free energy $\beta \mathcal{A}_k$ and the  direct correlation functions are obtained in Sections\ (\ref{flow-A}) and\ (\ref{flow-direct}) respectively. The derivation of the RG flow given here is elementary and does not call for the sophisticated mathematical apparatus developed for and applied in, deriving the HRT hierarchy. By the way, the genuine HRT  flow is obtained as a singular limit of our own smooth cut-off NPRG equations in section\ (\ref{HHRT}) where we point out that some the HRT flow equations for the direct correlation functions are ill-defined.
Section\ (\ref{RG-KSSHE}) is devoted to a transposition of the results of Section\ (\ref{RG-GC}) in the framework of the KSSHE theory of liquids. This scalar field theory of liquids  is obtained after performing a KSSHE transformation\cite{Kac,Siegert,Strato,Hubbard1,Hubbard2,Edwards} of the GC Boltzmann's factor and was studied intensively these past few years\cite{Brilliantov,Song,Cai-Mol,Cai-Pat1,Cai-Pat2}. KSSHE theory is a non-canonical field theory in the sense that the coupling between the source (the chemical potential for liquids) and the field is non linear which makes a direct comparison with Wetterich NPRG difficult. A mapping of KSSHE theory (and thus of the GC formalism) with a canonical field theory is however possible and discussed at length in Section\ (\ref{RG-canonic}). Our key result is that the flows of the average affective action $\Gamma_k$ of the canonical theory (governed by Wetterich equations) and that of the free energy $\beta \mathcal{A}_k$ (governed by the equations obtained in Section\ (II)) coincide (cf Section\ (\ref{lien}). This mapping makes possible the transposition of  all known-properties of the NPRG flows for canonical scalar field theories to the theory of liquids. An important consequence of this property is that our smooth cut-off NPRG theory for liquids yields a non-vanishing compressibility along the coexistence curve and is thus free of the shortcomings of HRT along  subcritical isotherms (see Section (\ref{comments})).  As an illustration of the issues of Sections\ (II,III,IV) we study the case of the Kac model\cite{Hansen,Kac,Kac1,Kac2,Kac3,Kac4} in Section\ (V). In this case the flow can be solved exactly by two complementary methods. We conclude in Section\ (VI).
\section{\label{RG-GC}The RG in the Grand-Canonical ensemble}
\subsection{\label{Prolego}Prolegomena} 
\subsubsection{\label{GC}The model}
We consider a simple fluid made  of identical hard spheres (HS) of diameter $\sigma$ with additional isotropic pair interactions $v(r_{ij})$ ($r_{ij}=| x_i -x_j |$, $x_i$  position of particle "$i$"). Since $v(r)$ is an arbitrary function of $r$  in the core, i.e. for $r \leq \sigma$, one can assume that $v(r)$ has been regularized in the core in such a way that  its Fourier transform $\widetilde{v}_{q}$ is a well behaved function of $q$ and that $v(0)$ is a finite quantity. 
We denote by $\Omega\subset\R^D$ ($D$ dimension of space) the domain occupied by the molecules of the fluid. For convenience $\Omega$ is chosen to be a cube of side $L$ and periodic boundary (PB) conditions are imposed so that the volume of $\Omega$ is $V=L^D$.

The fluid is at equilibrium in the grand canonical (GC) ensemble, $\beta=1/k_{\mathrm{B}}T$ is the inverse temperature ($k_{\mathrm{B}}$ Boltzmann's constant), and $\mu$  the chemical potential. In addition the particles are subject to an external potential $\psi(x)$ and we will denote by $\nu(x)=\beta (\mu-\psi(x))$ the dimensionless local chemical potential. We stick to notations usually adopted in standard textbooks devoted to the theory of liquids (see e.g. \cite{Hansen}) and  thus denote by $w(r)=-\beta v(r)$  \emph{minus } the dimensionless pair interaction. Moreover we  restrict ourselves to the case of attractive interactions, i.e. such that $\widetilde{w}(q)>0$ for all $q$.

In a given GC configuration $\mathcal{C}\equiv(N;x_1 \ldots x_N)$ the microscopic density of particles at point $x$  reads
\begin{equation}
\label{dens}
\widehat{\rho}(x|\mathcal{C}) =
\sum_{i=1}^{N} \delta^{(D)}(x-x_i) \; ,
\end{equation}
and the grand canonical partition function (GCPF)  $\Xi\left[ \nu \right] $ which encodes all the physics of the model at equilibrium is defined  as \cite{Hansen}
\begin{eqnarray}
\label{csi}\Xi\left[ \nu \right] &=&
\mathrm{Tr}\left[ \; \exp\left( -\beta \mathcal{H}_{\mathrm{GC}}\left[\mathcal{C}\right] 
\right) \right] \; , \nonumber \\
-\beta \mathcal{H}_{\mathrm{GC}}&=& -\beta
V_{\mathrm{HS}}+\frac{1}{2} \left\langle
\widehat{\rho}\vert w \vert\widehat{\rho} \right\rangle +
\left\langle  \overline{\nu}|\widehat{\rho} \right\rangle  \nonumber  \; ,\\ 
 \mathrm{Tr}\left[  \ldots \right] &=&
 \sum_{N=0}^{\infty}
\frac{1}{N!} \int_{\Omega}d1 \ldots dn \ldots \; ,
\end{eqnarray}
where $i \equiv x_i $ and $di\equiv d^{D}x_i$. For a given volume $V$ and a given inverse temperature $\beta$, $\Xi\left[ \nu \right]$ may be considered as a  functional of the local chemical potential $\nu(x)$  \cite{Hansen} which we have stressed by using a bracket.
In equation\ (\ref{csi}) $\beta V_{\mathrm{HS}}\left[ \mathcal{C}\right] $ denotes 
the HS contribution to the configurational energy and $\overline{\nu}=\nu+\nu_S$ where $\nu_S= - w(0)/2$ is $\beta$ times   the  self-energy of a
particle. We have  employed here convenient  brac-kets notations
\begin{eqnarray}
\left\langle  \overline{\nu}|\widehat{\rho} \right\rangle
 &\equiv&  \int_{\Omega} d1 \; \overline{\nu}(1)\widehat{\rho}(1) \\
\left< \widehat{\rho} \vert w \vert\widehat{\rho} \right> & \equiv &
\int_{\Omega} d1 d2\;
\widehat{\rho}(1|\mathcal{C})
 w(1,2)  \widehat{\rho}(2|\mathcal{C}) \; ,
\end{eqnarray}
but, at some places, we will also make use of matricial notations, e.g. \mbox{$\left< \widehat{\rho} \vert w \vert\widehat{\rho} \right>\equiv \widehat{\rho}(1) \cdot
w(1,2) \cdot \widehat{\rho}(2) $}.

Finally thermal averages of configurational quantities $\mathcal{A}\left[ \mathcal{C}\right]$ in the GC ensemble will be noted
\begin{equation}
\left\langle \mathcal{A}\left[ \mathcal{C}\right] \right\rangle_{\mathrm{GC}} = 
\frac{\mathrm{Tr}\left[ \mathcal{A}\left[ \mathcal{C}\right] \exp\left( -\beta
\mathcal{H}_{\mathrm{GC}}\left[ \mathcal{C}\right]\right) \right] }{\mathrm{Tr}\left[ \exp\left( -\beta
\mathcal{H}_{\mathrm{GC}}\left[ \mathcal{C}\right]\right) \right]} \; .
\end{equation}
\subsubsection{\label{GC-corre}Correlation functions}
As a consequence of a long history, the zoology of correlation functions in the theory of liquids is quite involved and two families of ordinary and truncated (or connected) density correlation functions are considered in the literature \cite{Hansen,Stell1,Stell2}.
The first family  is defined by means of functional derivatives of the GCPF
with respect to chemical potentials:
\begin{eqnarray}
\label{defcorre}
G^{(n)}[\nu](1, \ldots, n) &=&\left< \prod_{1=1}^{n} \widehat{\rho}
    (x_{i}  \vert \mathcal{C}) \right>_{\mathrm{GC}}
=\frac{1}{\Xi[\nu]}\frac{\delta^{n} \;\Xi[\nu]}
{\delta \nu(1) \ldots \delta \nu(n)}           \; ,\nonumber \\
G^{(n), T}[\nu](1, \ldots, n) &=&  \frac{\delta^{n} \log \Xi[\nu]}
{\delta \nu(1) \ldots \delta \nu(n)} \; .
\end{eqnarray}
Our notation emphasizes the fact that the  $G^{(n)}[\nu](1, \ldots, n)$ are functionals of the local chemical potential
$\nu(x)$ and functions of the coordinates $(1,\ldots, n) \equiv (x_{1},\ldots,
x_{n})$.
As it is well known we have \cite{Stell1,Stell2}
\begin{equation}
\label{iso}
G^{(n), T}[\nu](1,\ldots,n)= G^{(n)}[\nu]( 1,\ldots,n)
- \sum \prod_{m<n}G^{(m), T} [\nu](i_{1},\ldots,i_{m})  \; ,
\end{equation}
where the sum of products is carried out over all possible partitions of
the set $(1,\ldots,n)$ into subsets $(i_1,\ldots,i_m)$ of cardinal $m<n$. Of course $\rho[\nu](x) \equiv
G^{(n=1)}[\nu](x)=G^{(n=1), T}[\nu](x)$ is the local macroscopic density of the fluid.

The second family of correlation functions is obtained by considering derivatives of the GCPF with respect to the activity $z(i)=\exp(\nu(i))$. These objects will be denoted here by the symbol $\rho^{(n)}$ and are defined as
\begin{eqnarray}
\label{defcorre_bis}
\rho^{(n)}[\nu](1, \ldots, n) &=&\left< 
\sum_{i_1 \ne 1}^{N} \ldots \sum_{i_n \ne n}^{N} \delta^{(D)} (x_{i_1}-x_1) \ldots \delta^{(D)} (x_{i_n}-x_n)
\right>_{\mathrm{GC}} \; , \nonumber \\
&=& \frac{\prod_{i=1}^n\; z(i) }{\Xi[\nu]}\frac{\delta^{n} \;\Xi[\nu]}
{\delta z(1) \ldots \delta z(n)}           \; ,\nonumber \\
\rho^{(n), T}[\nu](1, \ldots, n) &=&  
 \prod_{i=1}^n z(i)  \; 
\frac{\delta^{n} \log \Xi[\nu]}
{\delta z(1) \ldots \delta z(n)} \; .
\end{eqnarray}
The ordinary $\rho^{(n)}$ and their connected parts $\rho^{(n), T}$ are related by  relations similar to\ (\ref{iso}). Of course $\rho^{(n=1)}\equiv G^{(n=1)}$ but, at higher orders,  $\rho^{(n)}$ and  $G^{(n)}$ differ by a host of  terms involving products of $\delta$ functions and local densities. We need to consider here only the case $n=2$ for which we have however the simple relation
\begin{eqnarray}
\label{n=2}
G^{(2)}(1,2)&=& \rho^{(2)}(1,2) + \rho(1) \delta(1,2)= \rho(1)\rho(2)
g(1,2)+ \rho(1) \delta(1,2) \nonumber \; , \\
G^{(2),T}(1,2)&=& \rho(1)\rho(2)h(1,2) + \rho(1) \delta(1,2) \; ,
\end{eqnarray}
where $g(1,2)$ ($\equiv g(r_{12})$ for a homogeneous system) is the usual pair correlation function of the theory of liquids, and $h(r)=g(r)-1$ tends to zero as $r \to \infty$.
\subsubsection{\label{GC-free}The GC free energy and the direct correlation functions.}
We review here some properties of the GC free energy and the direct correlation functions which will be used in the next sections.

It can be proved from first principles that the grand potential $\ln \Xi\left[ \nu\right]$ is a convex functional of the local chemical potential $\nu(x)$ at given $\beta$ and $V$ \cite{Goldenfeld,Cai-dft}. As a consequence of convexity  $\ln \Xi\left[ \nu\right]$ is continuous, its functional derivatives exist and are also continuous and it admits a Legendre transform $\beta \mathcal{A}\left[\rho \right]$ with respect to $\nu$ defined as
\begin{eqnarray}
\label{A-leg}
\beta \mathcal{A}\left[ \rho\right] &=& \sup_{\nu \in \mathcal{N}}\left\lbrace 
\left\langle \rho \vert \nu \right\rangle -\ln \Xi\left[\nu \right]  \right\rbrace  \; .
\end{eqnarray}
Although $\ln \Xi\left[ \nu\right]$ can be defined for any well behaved functions $\nu(x)$  (i.e. with some pedantry for $\nu \in \mathcal{N}$) the supremum in the left hand side of equation\ (\ref{A-leg}) exists only for a restricted class $\mathcal{R}$ of functions $\rho(x)$ (for instance if, for some $x \in \Omega$ one has $\rho(x) <0$ then $\rho \notin \mathcal{R}$). However if the supremum  exists then it is unique and is realized for a unique  $\nu^{\star}(x)$ given by the solution of the implicit equation: 
\begin{equation}
\label{nustar}
\rho(1) = \left.\frac{\delta\ln \Xi\left[ \nu\right]}{\delta \nu(1)}\right\vert_{\nu=\nu^{\star}} \; .
\end{equation}
$\beta \mathcal{A}\left[\rho \right]$ is then itself a convex functional of the local density $\rho(x)$ on the convex set $\mathcal{R}$ and is named the GC or Kohn-Sham free energy. Note that for a homogeneous system $\beta \mathcal{A}\left[\rho \right]$ and the usual Helmholtz free energy $\beta F(\rho)$ of the fluid coincide only in the thermodynamic limit. For instance, at temperatures below the critical temperature of the liquid-vapor transition
and at a finite volume $V$,  $\beta F(\rho)$ will exhibit a van der Waals loop while $\beta \mathcal{A}\left[\rho \right]$ will be convex (more precisely $\beta \mathcal{A}\left[\rho \right]$ is then the convex envelop of $\beta F(\rho)$)\cite{Goldenfeld,Cai-dft}. 
Since the Legendre transform is involutive we have for all $\nu \in \mathcal{N}$
\begin{subequations}
\begin{eqnarray}
\label{Xi-leg}
\ln \Xi\left[\nu \right]&=& \sup_{\rho \in \mathcal{R}}\left\lbrace 
\left\langle \nu \vert \rho \right\rangle - \beta \mathcal{A}\left[ \rho\right] \right\rbrace =
\left\langle \rho^{\star}\vert \nu \right\rangle -\beta \mathcal{A}\left[ \rho^{\star}\right] 
\; , \\ \label{statio}
\nu(1)&=& \left. \frac{\delta\beta \mathcal{A}\left[ \rho\right] }{\delta \rho(1)} \right\vert_{\rho=\rho^{\star}} \; .
\end{eqnarray}
\end{subequations}
Let us suppose that $\ln \Xi\left[ \nu \right]$ and  $\beta \mathcal{A}\left[\rho \right]$ depend in addition on some external parameter, say a scalar field $\alpha(x)$. It follows then from the stationary conditions\ (\ref{nustar}) and \ (\ref{statio}) that \cite{Zinn}
\begin{equation}
\label{utile}
\left. \frac{\delta \ln \Xi\left[\nu \right] }{\delta \alpha(1)}\right \vert_{\nu} = - 
\left. \frac{\delta \beta \mathcal{A}\left[\rho \right] }{\delta \alpha(1)}\right \vert_{\rho} \; ,
\end{equation}
where the partial functional derivatives of $\ln \Xi\left[\nu \right]$ and 
$\beta \mathcal{A}\left[\rho \right]$ with respect to $\alpha(x)$ must be taken at fixed $\nu$ and $\rho$  respectively; of course equation\ (\ref{utile}) is valid only if $\nu$ and $\rho$ are related by one of the two stationary conditions\ (\ref{nustar}) or \ (\ref{statio}).

It was shown in ref.\cite{Cai-Mol} that in the case of an attractive interaction (i.e.
$\widetilde{w}(q)>0$ for all $q$) we have the rigorous and useful  inequality
\begin{eqnarray}
\label{ineq}
\beta \mathcal{A}\left[ \rho\right] &\leq& \beta \mathcal{A}_{\mathrm{MF}}\left[ \rho\right] \; , \nonumber \\
\beta \mathcal{A}_{\mathrm{MF}}\left[ \rho\right] &=& \beta \mathcal{A}_{\mathrm{HS}}\left[ \rho\right] -\frac{1}{2} 
\left\langle \rho \vert w \vert \rho \right\rangle - \left\langle \rho
\vert \nu_{\mathrm{S}} \right\rangle \; .
\end{eqnarray}
Equation\ (\ref{ineq}) is valid for any  $\rho\in \mathcal{R}$ and $\mathcal{A}_{\mathrm{HS}}\left[ \rho\right]$ is the GC free energy of the HS reference system. The subscript "MF" which stands for "mean-field" emphasizes that $ \mathcal{A}_{\mathrm{MF}}$ is nothing but the van de Waals free energy\cite{Hansen}. Note that $\beta \mathcal{A}_{\mathrm{MF}}\left[ \rho \right]$ is \textit{a priori}  non convex, notably  in the two-phase region.

The n-point direct correlation function is defined in this paper as
\begin{equation}
\label{C}
C^{(n)}\left[\rho \right](1, \ldots,n)  = -
\frac{\delta \beta \mathcal{A}\left[\rho \right]}{\delta \rho(1) \ldots \delta \rho(n)} \; .
\end{equation}
The definition \ (\ref{C}) of $C^{(n)}$ includes an ideal gas contribution which is sometimes discarded explicitely by introducing a small cap $c^{(n)}$ defined as\cite{Stell1,Stell2}
\begin{equation}
\label{c}
C^{(n)}(1, \ldots,n)  = c^{(n)}(1, \ldots,n) + \frac{(-1)^n}{\rho(1)^{n-1}}
\prod_{i=2}^{n}\delta(1,i) \; .
\end{equation}
It follows from the properties of the Legendre transform that the linear operators $G^{(2), T}$ and $-C^{(2)}$ are positive definite and the inverse of each other:
\begin{equation}
G^{(2), T}(1,3)\; . \; C^{(2)}(3,2) = -\delta(1,2) \; .
\end{equation}
The above relation, when re-expressed in terms of $h(1,2)$ and $c(1,2) \equiv c^{(2)}(1,2)$ coincides with the well known Ornstein-Zernicke (OZ) equation of the theory of liquids \cite{Hansen,Stell1,Stell2}
\begin{equation}
h(1,2)=c(1,2) + c(1,3)\; . \; \rho(3)\; .\; h(3,2) \; .
\end{equation}
\subsection{\label{flow-A}Renormalization group flow for the effective energy} 
\subsubsection{\label{basics} Basic assumptions}
The trace operation which enters the definition\ (\ref{csi}) of the GCPF $\Xi\left[ \nu \right]$ may be seen as a sum over all the modes (or collective variables) $\widetilde{\rho}_{q}$, i.e. the Fourier coefficients of the microscopic density  $\widehat{\rho}(x)$ (cf equation\ (\ref{dens})). These modes are coupled via the HS interactions and the pair potential $\widetilde{w}(q)$ as well. Following the ideas which are the corner stone of the HRT theory\cite{HRT} we build a one-parameter family of models, called k-systems, which are indexed by a scale $k$ in momentum space. The k-system is a HS fluid with  pair interactions $\widetilde{w}_{k}(q)$ which couples the fast modes $\widetilde{\rho}_{q}$ ($q>k$) as in the original model  but  not  the slow modes ($q<k$). Therefore we ask that $\widetilde{w}_{k}(q) \simeq \widetilde{w}(q)$ for $q>k$ and $\widetilde{w}_{k}(q) \ll \widetilde{w}(q)$ for $q<k$.
Thus, as $k$ decreases, more and more density fluctuations $\widetilde{\rho}_{q}$ are integrated out. Denoting provisionally by $\Xi_{k}\left[\nu \right]$ and by $\beta \mathcal{A}_{k}\left[ \rho \right]$ the grand potential and the GC free energy of the k-system respectively we require that
\begin{eqnarray}
\label{limit-flow}
\lim_{k \to 0}\beta \mathcal{A}_{k}\left[ \rho \right]&=& \beta \mathcal{A}\left[ \rho \right] \; , \nonumber \\
\lim_{k \to \infty}\beta \mathcal{A}_{k}\left[ \rho \right]&=& \beta \mathcal{A_{\mathrm{MF}}}\left[ \rho \right] \; ,
\end{eqnarray}
where the MF Kohn-Sham free energy $\beta\mathcal{A_{\mathrm{MF}}}\left[ \rho \right]$ has been defined in equation\ (\ref{ineq}).
Note that the limit $k \to \infty $ in equation\ (\ref{limit-flow}) could be problematic and, in practice,  an ultra violet (UV) cut-off $k_{max} \simeq 1/\sigma$ on vectors $k$ can be introduced if needed.

In order to implement these ideas let us define
\begin{equation}
\label{wk}
\widetilde{w}_{k}(q)=\frac{\widetilde{w}(q)}{1 + \widetilde{w}(q)
\widetilde{R}_{k}(q)} \; ,
\end{equation}
where the regulator $\widetilde{R}_{k}(q)$ acts as a smooth IR cut-off for the pair potential and the  free energy (i.e it suppresses the contributions of the  slow modes $\widetilde{\rho}_{q}$, $q<k$ to $\beta \mathcal{A}_k\left[ \rho \right]$ ). Any function of the type $\widetilde{R}_{k}(q)= k^2 (1-\Theta_{\epsilon}(q,k))$ where  $\Theta_{\epsilon}(q,k)$ is some smooth version of the step function $\Theta(q-k)$ with width $\approx \epsilon$ will obviously do the job. More precisely one demands that $\Theta_{\epsilon}(q,k)\simeq 1$ for $q \geq  k+ \epsilon$ and
$\Theta_{\epsilon}(q,k)\simeq 0$ for $q \leq k- \epsilon$. One could think of  $\Theta_{\epsilon}(q,k)= \left( 1+ \tanh\left((q-k)/\epsilon \right) \right) /2$  for instance. The choice of an analytical form for $\widetilde{R}_{k}(q)$ should be unimportant in principle but must be considered with some care when approximations are introduced at a further stage.
Let us summarize the behavior of $\widetilde{R}_{k}(q)$ in the various regimes of interest:
\begin{subequations}
\label{111}
\begin{eqnarray}
\label{1a}
\widetilde{R}_{k}(q)& \geq& 0 \; \; \forall q \; , \\
\label{1b}
\widetilde{R}_{k}(q)& \approx & \; k^2 \; \; \mathrm{ for } \; \; q \ll k \; , \\ 
\label{1c}
\widetilde{R}_{k}(q) & \to  & 0  \; \; \mathrm{fastly \;  for } \; \; q \to \infty \; .
\end{eqnarray}
\end{subequations}
The positivity condition\ (\ref{1a}) ensures that, at scale-k the pair potential $w_k(r)$ is attractive, but "less" than $w(r)$, since one has $0\leq\widetilde{w}_{k}(q)\leq \widetilde{w}(q) \; \forall q$. Indeed, as required $\widetilde{w}_{k}(q)\approx \widetilde{w}(q) \; \mathrm{ for}\; q>k$ and  $\widetilde{w}_{k}(q)\approx 0 \; \mathrm{ for}\; q<k $ are consequences of equations\ (\ref{1b}) and\ (\ref{1c}). Finally we note that
\begin{subequations}
\begin{eqnarray}
\lim_{k\to 0 }\widetilde{w}_{k}(q)& =& \widetilde{w}(q) \; , \\
\lim_{k\to \infty }\widetilde{w}_{k}(q)& =& 0 \; ,
\end{eqnarray}
\end{subequations}
hence, when the flow starts at $k=\infty$ one deals with a fluid of pure hard spheres and, when it stops at $k=0$ one recovers the full interacting system.

Parola has proposed his own smooth cut-off version of HRT equations\cite{Parola}, however his choice for $w_k$, i.e. $\widetilde{w}_k(q)= \widetilde{w}(q) -k^{2-\eta}\widetilde{w}(q/k) $ ($\eta>0$ Fisher's exponent) does not enter the class of potentials defined at equation\ (\ref{wk}).

$\Xi_{k}\left[ \nu \right]$ will be defined according to equation\ (\ref{csi}) with the replacement $w \rightarrow w_k$. Since it is the GCPF of a HS fluid with additional pair interactions $w_k(r)$,  $\ln\Xi_{k}\left[ \nu \right]$ is a convex functional of the local potential $\nu(x)$. Denoting by $\beta \mathcal{A}_{k}^{'}\left[ \rho \right] $ its Legendre transform, one has the usual couple of relations
\begin{eqnarray}
\ln \Xi_k\left[\nu \right]&=& \sup_{\rho \in \mathcal{R}}\left\lbrace 
\left\langle \nu \vert \rho \right\rangle - \beta \mathcal{A}_k^{'}\left[ \rho \right] \right\rbrace \; \;  \forall \nu \in \mathcal{N}, \nonumber \\
\beta \mathcal{A}_k^{'}\left[ \rho\right] &=& \sup_{\nu \in \mathcal{N}}\left\lbrace 
\left\langle \rho \vert \nu \right\rangle -\ln \Xi_k\left[\nu \right]  \right\rbrace \; \; \forall \rho \in \mathcal{R} \; .
\end{eqnarray}
Recall that $\beta \mathcal{A}_k^{'}\left[ \rho\right]$ is then a convex functional of the density. However $\beta \mathcal{A}_k^{'}\left[ \rho\right]$
has not the required behavior\ (\ref{limit-flow}) for $k \to \infty$ and, following Reatto \textit{et al.} \cite{HRT}, one is led to introduce the  effective average free energy
\begin{equation}
\label{def-Ak}
\beta \mathcal{A}_k\left[ \rho\right]= \beta \mathcal{A}_k^{'}\left[ \rho\right] + \frac{1}{2}\; \left\langle \rho \vert w_{k} -w  \vert \rho \right\rangle \; +  \left\langle \nu_{\mathrm{S}, \; k} -\nu_S \vert \rho \right\rangle  \; ,
\end{equation}
where $\nu_{\mathrm{S}, \; k}=-w_k(0)/2$ is the self energy of particles of the k-system.
Now, for $k\rightarrow 0$ one has $w_k \rightarrow w \; \Rightarrow \Xi_k \rightarrow
\Xi \; \Rightarrow \beta \mathcal{A}_{k}\rightarrow \beta \mathcal{A}_{k=0}^{'}= \beta \mathcal{A}$. Moreover, for $k\rightarrow \infty$ one has $w_k \rightarrow 0 \; \Rightarrow \Xi_k \rightarrow
\Xi_{\mathrm{HS}} \; \Rightarrow \beta \mathcal{A}^{'}_{k}\rightarrow \beta \mathcal{A}_{\mathrm{HS}} \; \Rightarrow
\beta \mathcal{A}_{k}\rightarrow  \beta \mathcal{A}_{\mathrm{MF}} $ as required.

Note that since $\widetilde{w}_{k}(q) -\widetilde{w}(q)<0 ,\; \forall q$ then $\beta \mathcal{A}_k\left[ \rho\right]$ is the sum of a convex-up functional (i.e. $\mathcal{A}_k^{'}\left[ \rho\right]$) and a convex-down quadratic form, hence, in general, it will not be convex notably in the two phase region at $k \neq 0$ (in particular at $k=\infty$); convexity will be recovered only in the $k\to 0$ limit\cite{Wetterich}.

Before deriving the flow equations for $\mathcal{A}_k\left[ \rho\right]$ let us mention some important properties of this flow and precise its initial conditions at $k=\infty$. Firstly we note that it follows readily from equations\ (\ref{ineq}) and\ (\ref{def-Ak}) that for any scale-k (for $k=0$ we already know the result) and $\rho \in
\mathrm{R}$ one has 
\begin{equation}
\label{bound-k} \beta \mathcal{A}_k\left[ \rho\right]
\leq  \beta \mathcal{A}_{\mathrm{MF}}\left[ \rho\right] \equiv
\beta \mathcal{A}_{k=\infty}\left[ \rho\right] \; ,
\end{equation}
from which however it would be hazardous (and in fact wrong in general but true in the toy model of Section\ V) to infer that $\beta \mathcal{A}_{k}\left[ \rho\right]$ is a decreasing function of k for any density $\rho(x) \in \mathcal{R}$, even a uniform one. 

The "true" direct correlation functions at scale-k, noted $C^{' \; (n)}_{k}$, are of course defined according to equation\ (\ref{C}) with the replacement $\mathcal{A} \rightarrow \mathcal{A}_{k}^{'}$. In addition we also introduce
effective direct correlation functions as
\begin{equation}
\label{C-k}
C^{(n)}_{k}\left[\rho \right](1, \ldots,n)  = -
\frac{\delta \beta \mathcal{A}_{k}\left[\rho \right]}{\delta \rho(1) \ldots \delta \rho(n)} \; .
\end{equation}
It follows from equation\ (\ref{def-Ak}) that the $C^{(n)}_{k}$ and the
$C^{' \; (n)}_{k}$ coincide for $n\geq 3$ but that for $n=2$ one has
\begin{equation}
\label{utile-bis}
C^{' \; (2)}_{k}\left[\rho \right] (1,2)=
C^{(2)}_{k}\left[\rho \right] (1,2) \;  + \; w_k(1,2)\;  - \; w(1,2) \; ,
\end{equation}
from which we infer from the convexity of $\beta \mathcal{A}_{k}^{'}\left[ \rho\right]$ that, in k-space
\begin{equation}
\label{sign-den}
\widetilde{C}^{(2)}_{k}(q) \; + \; \widetilde{w}_k(q) \; - \; \widetilde{w}(q) < 0 \; \; \forall q \; .
\end{equation}
Finally, since $\beta \mathcal{A}_{k=\infty}=\beta \mathcal{A}_{\mathrm{MF}}$ the initial conditions for the $C^{(n)}_{k}$ are:
\begin{eqnarray}
\label{cond-init}
C^{(1)}_{k \to \infty}\left[\rho \right](1)&=& C^{(1)}_{\mathrm{HS}}\left[\rho \right](1) + w(1,2)\cdot \rho(2) + \nu_{\mathrm{S}} \; , \nonumber \\
C^{(2)}_{k \to \infty}\left[\rho \right](1,2) &=& 
C^{(2)}_{\mathrm{HS}}\left[\rho \right](1,2) +w(1,2) \; , \nonumber \\
C^{(n)}_{k \to \infty}\left[\rho \right](1,\ldots,n) &=& 
C^{(n)}_{\mathrm{HS}}\left[\rho \right](1,\ldots,n) \; \mathrm{ for} \; \; n \geq 3 \; .
\end{eqnarray}
Note that the initial condition for $C^{(2)}_{\infty}\left[\rho \right](1,2)$ coincides with the so-called random phase approximation (RPA) of the theory of liquids \cite{Hansen,Cai-Mol,HRT}. We will assume that the direct correlation functions of the HS fluid are exactly known which is of course not actually the case; however some excellent approximations for the $C^{(n)}_{\mathrm{HS}}$ can be found in the literature\cite{Hansen}.
\subsubsection{\label{flow-eq} Flow equations for $ \ln\Xi_{k}$ and $\beta \mathcal{A}_{k}$}
It follows readily from the definition\ (\ref{csi}) of the GCPF $\Xi_{k}\left[ \nu \right]$ (with $w \rightarrow w_k$) that the partial derivative $\partial_{k}$ of
$\ln \Xi_{k}\left[ \nu \right]$ with respect to scale-k (at fixed $\nu$) is given by
\begin{eqnarray}
\label{flow-csik}
\partial_{k} \ln \Xi_{k}\left[ \nu \right]&=& \frac{1}{2}
\left\langle \sum_{i\neq j} \partial_{k} w_{k}(i,j)\right\rangle_{\mathrm{GC}} \nonumber \\ 
&=& \frac{1}{2} \int d1 \; d2\; \partial_{k} w_{k}(1,2) \;  \rho^{(2)}_{k}\left[\nu \right] (1,2) \; ,
\end{eqnarray}
where $\rho^{(2)}_{k}[\nu](1,2)$ is the pair correlation of the k-system as defined in equation\ (\ref{defcorre_bis}). From the stationary property\ (\ref{utile}) with $\alpha(x)\equiv k$ we deduce that
\begin{eqnarray}
\label{flowAprime}
\partial_{k} \beta \mathcal{A}_{k}^{'}\left[ \rho \right] &=& - \frac{1}{2} \int d1 \; d2\; \partial_{k} w_{k}(1,2) \;  \rho^{(2)}_{k}\left[\nu \right](1,2) \; ,
\end{eqnarray}
where $\rho$ and $\nu$ are related through the equations\ (\ref{nustar}) and\ (\ref{statio}). 
Injecting now equation\ (\ref{flowAprime}) in the definition\ (\ref{def-Ak}) yields the flow equation for the effective free energy, i.e.
\begin{eqnarray}
\label{flowA}
\partial_{k} \beta \mathcal{A}_{k}\left[ \rho \right]&=& - \frac{1}{2} \int d1 \; d2\; \partial_{k} w_{k}(1,2) \;  G^{(2) \; T}_{k}\left[\nu \right](1,2) \; ,
\end{eqnarray}
where we have made use of equation\ (\ref{n=2}). Note that  equation\ (\ref{flowA}) is somehow reminiscent, at least formally, of the charging process of the theory of classical fluids\cite{Hill}. Recall that $G^{(2) \; T}_{k}$ is minus the inverse of $C^{' \; (2)}_{k}$ and, taking into account the relation\ (\ref{utile-bis}), we get the flow equation
\begin{eqnarray}
\label{flowA-final}
\partial_{k} \beta \mathcal{A}_{k}\left[ \rho \right]&=&  \frac{1}{2} \int d1 \; d2\; \partial_{k} w_{k}(1,2) \; \left\lbrace 
C_{k}^{(2)}\left[\rho \right] \; + \; w_k \; -w 
\right\rbrace^{-1} (1,2) \; .
\end{eqnarray}
This beautiful equation which was obtained in a trivial way is exact and thus horribly complicated. Mathematically
it is a functional partial derivative equation (both $\beta \mathcal{A}_{k}$ and
$C_{k}^{(2)}$ are functionals of $\rho(x)$), needless to say impossible to solve exactly. But two types of approximations have been be proposed  to solve it, at least numerically; they will be briefly evoked in Section VI.

The one-loop structure of the flow equation\ (\ref{flowA-final}) should be more easily comprehended if  rewritten as
\begin{eqnarray}
\label{flowA-final-bis}
\partial_{k} \beta \mathcal{A}_{k}&=& - \frac{1}{2} \; 
\widetilde{\partial}_k  \mathrm{Tr} \ln\left(-\; C_{k}^{(2)} \; - \; w_k \; +w\right)  \; ,
\end{eqnarray}
where $\widetilde{\partial}_k \ldots =\partial_k w_k \; \partial \ldots/\partial w_k$ acts only on the k-dependence of $w_k$ and not on $C_{k}^{(2)}$ and the trace operation is understood in the usual matricial sense, i.e. as an integral over the position variables $1$ and $2$. Note that the signs have been chosen in such a way that the argument of the $\ln$ is positive definite. It is the place to mention that it is found that, in  KSSHE or CV field theories, the one-loop approximation for the "true" GC free energy $\beta \mathcal{A}_{k}^{'}$ of the k-system reads as \cite{Cai-Mol,Cai-Pat1,Cai-Pat2}
\begin{eqnarray}
\label{1-loop}
\beta \mathcal{A}_{k}^{' \; (\mathrm{1-loop})}\left[\rho \right]  &=&\beta \mathcal{A}_{\mathrm{MF}, \;k}\left[\rho \right] +\frac{1}{2}
\mathrm{Tr}\ln \left( 1-w_k\cdot G_{\mathrm{HS}}^{(2), \; T}\right) \; , 
\end{eqnarray}
where $\beta \mathcal{A}_{\mathrm{MF}, \;k}\left[\rho \right]$ is given by equation\ (\ref{ineq}) with  $w \rightarrow w_k$. Note that the expression\ (\ref{1-loop}) of $\beta \mathcal{A}_{k}^{'}$ coincides with that obtained in the  RPA approximation of the theory of liquids\cite{Hansen,HRT}.
Taking into account equation\ (\ref{def-Ak}) one finds for the average free energy
\begin{eqnarray}
\label{1-loop-b}
\beta \mathcal{A}_{k}^{(\mathrm{1-loop})}\left[\rho \right]  &=&\beta \mathcal{A}_{\mathrm{MF}}\left[\rho \right] +\frac{1}{2}
\mathrm{Tr}\ln \left( \left(\left( \leftrightarrows C_{\mathrm{HS}}^{(2)}+w\right)  + w_k - w\right)\cdot C_{\mathrm{HS}}^{(2) \; \;-1} \right)  \; , 
\end{eqnarray}
where we have made use of the OZ equation $C_{\mathrm{HS}}^{(2)\; -1}= - G_{\mathrm{HS}}^{(2)\; T}$ for the reference HS fluid. Thus, substituting $ C_{\mathrm{HS}}^{(2)}+w$ (i.e. the RPA approximation for  $C_{k}^{(2)}$ since $C^{' \; (2)}_{k \;\mathrm{(RPA)}}=C_{\mathrm{HS}}^{(2)}+w_k \Rightarrow  C^{(2)}_{k \;\mathrm{(RPA)}}=C_{\mathrm{HS}}^{(2)}+w $) by the full $C_{k}^{(2)}$ turns the one-loop result into an exact one !

For a homogeneous system $\beta \mathcal{A}_{k}\left[ \rho \right]=V f_k(\rho)$ where $\beta$ times the free energy per unit volume $f_k(\rho)$ is a function (not a functional) of $\rho$ and the flow equation for $f_k(\rho)$ reads as
\begin{eqnarray}
\label{flow-f}
\partial_{k}f_k &=&
 \frac{1}{2} \int_{q} \frac{\partial_{k} \widetilde{w}_{k}(q)
}{\widetilde{C}_{k}^{(2)} (q) + \widetilde{w}_{k}(q) - \widetilde{w}(q)} \; ,
\end{eqnarray}
where $\int_q \equiv \int d^{D}q/(2 \pi)^D$ and $\widetilde{C}_{k}^{(2)} (q)$ is the Fourier transform of $C_k^{(2)}(r)$. Note that the right hand side is non singular since the denominator is negative definite (cf equation\ (\ref{sign-den})). It is a general property that the flow of $f_k$ has neither UV or IR singularities, in particular IR singularities (near a critical point) are smoothened by k and they build up progressively as the scale-k is lowered\cite{Wetterich,Delamotte}.

One can deduce from equation\ (\ref{flowA-final}) an infinite tower of equations for  the $C^{(n)}_{k}\left[\rho \right](1, \ldots,n)$. This is the subject of next section\ (\ref{flow-direct}).
\subsection{\label{flow-direct} RG flow of the effective direct correlation functions} 
\subsubsection{\label{real} Real space}
The RG flow for the  effective direct correlation function $C_k^{(n)}$ is obtained by taking $n$ successive functional derivatives of both members of equation\ (\ref{flowA-final}) with respect to the density $\rho(x)$.
Let us first work out the case $n=1$ explicitely. Since $(C_k^{(2)}-w + w_k)^{-1} =-G_k^{(2),  \; T}$ one has, by taking the functional derivative of equation\ (\ref{flowA-final}) with respect to $\rho(1)$
\begin{equation}
\partial_k C_k^{(1)}(1)=
\frac{1}{2} \; \partial_k w_k(1^{'},2^{'})\frac{ \delta G_k^{(2),  \; T}(1^{'},2^{'})}{\delta \rho(1)} \; .
\end{equation}
It follows from the general result $\delta A^{-1}(1,2)= -A^{-1}(1,1^{'})\cdot\delta A(1^{'},2^{'})\cdot A^{-1}(2^{'},2)$ valid for any linear operator $A(1,2)$ and variation $\delta$ that
\begin{equation}
\label{rule-vertex}\frac{ \delta G_k^{(2),  \; T}(1,2)}{\delta \rho(3)}=G_k^{(2),  \; T}(1,1^{'})\cdot C_k^{(3)}(1^{'},2^{'},3) \cdot G_k^{(2),  \; T}(2^{'},2) \; ,
\end{equation}
yielding readily
\begin{equation}
\label{flow-C1}
\partial_k C_k^{(1)}(1)=\frac{1}{2} \; \partial_k w_k(1^{'},2^{'}) \cdot G_k^{(2),  \; T}(1^{'},1^{''}) \cdot 
C_k^{(3)}(1^{''},2^{''},1) \cdot G_k^{(2),  \; T}(2^{''},2^{'}) \; . 
\end{equation}
The flow equation for $C^{(n=2)}$ can be obtained by taking the functional derivatives of equation\ (\ref{flow-C1}) with respect to  $\rho(2)$. But we will not do it in that way. In fact,
the structure of the hierarchy of the flow equations for the $C^{(n)}$ is more easy to grasp on the basis of a diagrammatic analysis and equation\ (\ref{flow-C1}) will be rederived now by using Feynman diagrams and simple diagrammatic rules.  These diagrams will be built out from two kinds of vertex
\begin{subequations}
\begin{eqnarray}
\begin{picture}(55,35)(0,7)
\DashLine(0,10)(20,10){3}
\BBoxc(23,10)(6,6)
\DashLine(26,10)(46,10){3}
\Text(0,0)[]{1}  \Text(46,0)[]{2}
\end{picture} 
&=& \; \; \partial_k w_k(1,2)  \; , \\
\begin{picture}(55,35)(0,7)
\Vertex(20,10){3}
\DashLine(20,10)(5,20){3}
\DashLine(20,10)(5,0){3}
\DashLine(20,10)(35,20){3}
\DashLine(20,10)(35,0){3}
\Text(45,20)[]{$1$}
\Text(-5,20)[]{$2$}
\Text(45,0)[]{$n$}
\Text(-5,0)[]{\ldots}
\end{picture}& =& \; \; C_k^{(n)} (1,2, \ldots,n) \; \; (n \geq 1) \; \; ,
\end{eqnarray}\end{subequations}
and the propagator
\begin{equation}
\begin{picture}(55,35)(0,7)
\Line(0,10)(46,10)
\Text(0,0)[]{$1$}  \Text(46,0)[]{$2$}\end{picture} 
= \; \; G_k^{(2), \; T}(1,2) \; .
\end{equation}
We can now represent the flow equation\ (\ref{flowA-final}) diagrammatically as 
\begin{equation}
\label{flowA-diag}\partial_k \beta \mathcal{A}_k = - \frac{1}{2}
\begin{picture}(55,20)(0,7)
\CCirc(20,10){15}{Black}{White}
\BBoxc(5,10)(5,5)
\end{picture} \; .
\end{equation}
The action of a functional derivative with respect to $\rho(1)$ on the vertex $C_k^{(n)}$ and the propagator $G_k^{(2), \; T}$ are given by simple diagrammatic rules:
\begin{subequations}
\label{rules}
\begin{eqnarray}
\frac{\delta }{\delta \rho(n)} \begin{picture}(65,45)(0,7)
\Vertex(30,10){3}
\DashLine(30,10)(15,20){3}
\DashLine(30,10)(15,0){3}
\DashLine(30,10)(45,20){3}
\Text(50,25)[]{$1$}
\Text(15,-5)[]{$n$}
\Text(15,25)[]{\ldots}
                              \end{picture}
			      &=&\begin{picture}(65,45)(0,7)
\Vertex(30,10){3}
\DashLine(30,10)(15,20){3}
\DashLine(30,10)(15,0){3}
\DashLine(30,10)(45,20){3}
\DashLine(30,10)(45,0){3}
\Text(50,25)[]{$1$}
\Text(15,25)[]{\ldots}
\Text(50,-5)[]{$n+1$}
\Text(15,-5)[]{$n$}                             \end{picture} \; , \label{o1}\\
\frac{\delta }{\delta \rho(3)} \begin{picture}(65,45)(0,7)
\Line(15,10)(55,10)\Text(15,0)[]{$1$} \Text(55,0)[]{$2$} 
                               \end{picture}
			       &=&\begin{picture}(65,45)(0,7)
\Line(15,10)(55,10) \Text(15,0)[]{$1$} \Text(55,0)[]{2} \Vertex(35,10){3}
\DashLine(35,10)(35,30){3} \Text(45,30)[]{$3$} 	       
			       \end{picture} \; ,\label{o2}
\end{eqnarray}\end{subequations}
where equation\ (\ref{o1}) follows from the very definition of the direct correlation functions and equation\ (\ref{o2}) is the diagrammatic representation of  equation\ (\ref{rule-vertex}). Taking now the functional derivative of equation\ (\ref{flowA-diag}) with respect to $\rho(1)$ by taking into account the rules\ (\ref{rules}) one finds that
\begin{equation}
\label{cont1}\partial_k \begin{picture}(65,45)(0,7)
\DashLine(15,10)(55,10){3}\Vertex(55,10){3} \Text(10,10)[]{$1$}
           \end{picture}
	   = \; \; \frac{1}{2}\; \; \begin{picture}(55,20)(0,7)
\CCirc(20,10){15}{Black}{White}
\BBoxc(5,10)(5,5) \Vertex(35,10){3} \DashLine(35,10)(55,10){3}\Text(60,10)[]{$1$}
\end{picture} \; \; \; \; \; ,
\end{equation}
which is nothing but the diagrammatic representation of equation\ (\ref{flow-C1}). Continuating the process one finds
\begin{equation}
\label{cont2}\partial_k \begin{picture}(65,45)(0,7)
\DashLine(15,10)(55,10){3}\Vertex(35,10){3}\Text(10,10)[]{$1$}\Text(60,10)[]{$2$}
\end{picture}=
\; \; \frac{1}{2}\; \; \begin{picture}(55,20)(0,7)
\CCirc(20,10){15}{Black}{White}
\BBoxc(5,10)(5,5) \Vertex(35,10){3} \DashLine(35,10)(55,20){3}\Text(60,20)[]{$1$}
\DashLine(35,10)(55,0){3}\Text(60,0)[]{$2$} \end{picture}
\; \; \; \;  + \; \;
\begin{picture}(55,20)(0,7)
\CCirc(20,10){15}{Black}{White}
\BBoxc(5,10)(5,5) 
\Vertex(20,25){3} \DashLine(20,25)(20,40){3} \Text(25,40)[]{$1$}
\Vertex(20,-5){3} \DashLine(20,-5)(20,-20){3}\Text(25,-20)[]{$2$}
\end{picture} \; \; \; ,
\end{equation}
which can be translated algebraically by
\begin{eqnarray}
\partial_k C_k^{(2)}(1,2)&=&\frac{1}{2}\partial_k w(a,b)G_k^{(2), \; T}(b,c)
G_k^{(2), \; T}(a,d) C_k^{(4)}(c,d,1,2) \; \nonumber \\
&+& \partial_k w(a,b) G_k^{(2), \; T}(a,c) G_k^{(2), \; T}(b,d) \times \nonumber \\
&\times & C_k^{(3)}(c,e,2) G_k^{(2), \; T}(e,f) C_k^{(3)}(f,d,1)  \; .
\end{eqnarray}
Flow equations for $C_k^{(n)}$ of higher orders are obtained in the same vein by making use \textit{ad libitum} of the diagrammatic rules\ (\ref{rules}).
Some comments are in order.
\begin{itemize}
\item The equation for $\partial_k C_k^{(n)}$ involves \textit{inter alias} the vertex $C_k^{(n+1)}$ and  $C_k^{(n+2)}$, therefore the hierarchy never closes. Possible approximations consist in enforcing a closure at some order $n$\cite{HRT,Wetterich,Wschebor,Wschebor2}.
\item Contemplation of equations\ (\ref{cont1},\ \ref{cont2}) and a little thought reveal that the one-loop structure is present at each order $n$ and therefore only one  integral on internal variables survives. 
\item All the expressions  for the odd $\partial_k C_k^{(2n+1)}$ include diagrams with at least one odd vertex $C_k^{(2m+1)}$. Therefore if at any scale k (in practice $k=\infty$) all the odd $C_k^{(2m+1)}$ happen to vanish they will remain exactly zero at smaller scales k. This property is present only for peculiar models of fluids with $Z_2$ symmetry (for instance some versions of the lattice gas with special particle-hole symmetry), but not in general.
\end{itemize}
\subsubsection{\label{Fourier} Fourier space}
We restrict ourselves to homogeneous systems. Then, as usual, in momentum space, we factor out and evaluate the momentum conserving $\delta$ function so $n$-point correlation functions  $\widetilde{G}^{(n)}(q_1,\ldots,q_n)$ are defined only when $q_1 + \ldots +q_n=0$. More precisely one has for instance
\begin{eqnarray}
\widetilde{G}^{(n),\; T}(q_1,\ldots,q_n)& =& \widehat{\delta}(q_1+\ldots q_n)
\int_{x_1 \ldots x_n}\exp\left( i\left( q_1 x_1 +\ldots q_{n-1}x_{n-1}\right)\right) \times  \nonumber \\
&\times& G^{(n),\; T}(x_1,\ldots,x_{n-1},0) \; , 
\end{eqnarray}
where $\widehat{\delta}(q) \triangleq (2\pi)^{D}\delta^{D}(q)$ and $\int_x  \triangleq \int_{\Omega} d^D x$. In addition, in two-point functions, we solve $q_1=-q_2=q$ and recognize that they are functions only of $q=\vert \vec{q} \vert$, e.g. $\widetilde{G}^{(2),\; T}(q)$.
Feynman diagrams in Fourier space will be build from the vertex
\begin{subequations}
\begin{eqnarray}
\begin{picture}(55,35)(0,7)
\ArrowLine(0,10)(20,10)
\BBoxc(23,10)(6,6)
\ArrowLine(46,10)(26,10)
\Text(10,0)[]{$q$}  \Text(36,0)[]{$q$}
\end{picture} 
&=& \; \; \partial_k \widetilde{w}_k(q)  \; , \\
\begin{picture}(55,35)(0,7)
\Vertex(20,10){3}
\ArrowLine(5,20)(20,10)
\ArrowLine(5,0)(20,10)
\ArrowLine(35,20)(20,10)
\ArrowLine(35,0)(20,10)
\Text(45,20)[]{$q_1$}
\Text(-5,20)[]{$q_2$}
\Text(45,0)[]{$q_n$}
\Text(-5,0)[]{\ldots}
\end{picture}& =& \; \; \widetilde{C}_k^{(n)} (q_1,\ldots,q_n) \; \; (n \geq 1) \; \; ,
\end{eqnarray}\end{subequations}
and the propagator
\begin{equation}
\begin{picture}(55,35)(0,7)
\ArrowLine(0,10)(23,10)
\ArrowLine(46,10)(23,10)
\Text(10,20)[]{$q$} \Text(30,20)[]{$-q$}\end{picture} 
= \; \; \widetilde{G}_k^{(2), \; T}(q)\equiv \frac{-1}{\widetilde{C}_k^{(2)}(q) 
-\widetilde{w}(q) + \widetilde{w}_k(q)}\; .
\end{equation}
One has now for the  effective average free energy per unit volume (cf equation\ (\ref{flowA-diag}))
\begin{eqnarray}
\label{flowf-diag}\partial_k  f_k  & =& - \frac{1}{2}
\begin{picture}(55,40)(0,7)
\ArrowArc(20,10)(15,0,180)
\ArrowArcn(20,10)(15,360,180)
\Text(20,32)[]{$q$} \Text(20,-12)[]{$-q$} 
\BBoxc(5,10)(5,5)
\end{picture} \; = \; 
\frac{1}{2} \int_q \frac{\partial_k \widetilde{w}_k(q)}{\widetilde{C}^{(2)}_k(q) - \widetilde{w}(q)+\widetilde{w}_k(q)} \; , \\
\nonumber
\end{eqnarray}
One then easily obtains the flows for the low order effective direct correlation functions in momentum space. With the convention that the impulsion is conserved at each vertex and that all the diagrams drawn below are shorn of their external lines one obtains 

\begin{eqnarray}
\label{flow-C1k}
\partial_k \widetilde{C}^{(1)}(0) &=&
\partial_k\begin{picture}(65,45)(0,7)
\ArrowLine(15,10)(55,10)\Vertex(55,10){3} \Text(35,20)[]{$0$}
\end{picture} 
= \; \; \frac{1}{2}\; \; \begin{picture}(55,45)(0,7)
\ArrowLine(0,10)(30,10) \Text(15,20)[]{$0$}
\Vertex(30,10){3}	   
\ArrowArc(45,10)(15,90,180) \Text(35,30)[]{$q$}
\ArrowArcn(45,10)(15,270,180) \Text(35,-10)[]{-$q$}
\ArrowArcn(45,10)(15,90,0)   \Text(55,30)[]{-$q$}
\ArrowArc(45,10)(15,270,360)  \Text(55,-10)[]{$q$}
\BBoxc(60,10)(5,5) 
\end{picture} \nonumber \\ \nonumber \\
&=&\frac{1}{2}\int_q\widetilde{C}^{(3)}(0,q,-q)\widetilde{G}_k(q)^2 \partial_k \widetilde{w}_k(q) \; ,
\end{eqnarray}
and
\begin{eqnarray}
\label{flow-C2k}\partial_k \widetilde{C}^{(2)}(p) &=&\partial_k\begin{picture}(65,45)(0,7)
\ArrowLine(0,10)(30,10) \Text(15,20)[]{$p$}
\Vertex(30,10){3}
\ArrowLine(60,10)(30,10) \Text(45,20)[]{$-p$}
\end{picture}  \nonumber \\
&=&\frac{1}{2} \; \; \begin{picture}(65,45)(0,7)
\ArrowLine(0,0)(30,10) \Text(15,-5)[]{$p$}
\Vertex(30,10){3}
\ArrowLine(60,0)(30,10) \Text(45,-5)[]{$-p$}
\ArrowArcn(30,25)(15,180,90)
\ArrowArc(30,25)(15,0,90)
\ArrowArcn(30,25)(15,360,270)
\ArrowArc(30,25)(15,180,270)
\Text(50,35)[]{$q$}
\Text(50,15)[]{$-q$}
\Text(5,35)[]{$-q$}
\Text(10,15)[]{$q$}
\BBoxc(30,40)(5,5)
\end{picture}\;  +\; \; \; \; \; 
\begin{picture}(110,45)(0,7)
\ArrowLine(0,10)(30,10) \Text(-5,15)[]{$p$} \Text(25,20)[]{$q$}
\Vertex(30,10){3}
\Vertex(60,10){3}
\ArrowLine(90,10)(60,10)\Text(100,15)[]{$-p$} \Text(65,20)[]{$-q$} 
\ArrowArc(45,10)(15,45,90)
\ArrowArc(45,10)(15,135,180)
\ArrowArc(45,10)(15,270,360) \Text(62,-10)[]{$p+q$}
\ArrowArcn(45,10)(15,270,180)\Text(22,-10)[]{$-p-q$}
\ArrowArcn(45,10)(15,135,90)
\ArrowArcn(45,10)(15,45,0)
\BBoxc(45,25)(5,5) \Text(40,37)[]{$-q \; \; \; q$}
\end{picture} \nonumber \\
\nonumber \\
&=&\frac{1}{2}\int_q \widetilde{C}^{(4)}(p,q,-q,-p)\widetilde{G}^{(2), \; T}(q)^2 \partial_k  \widetilde{w}(q) \nonumber \\
&+& \int_q \widetilde{C}^{(3)}(p,q,-p-q)\widetilde{C}^{(3)}(-q,-p,p+q)
\widetilde{G}^{(2), \; T}(q)^2 \widetilde{G}^{(2), \; T}(p+q) \partial_k  \widetilde{w}(q) \; . 
\end{eqnarray}

\subsection{\label{HHRT} HRT as an ultra-sharp cut-off limit} 
In HRT  the pair potential of the k-system is defined as\cite{HRT}
\begin{eqnarray}
\label{wkHRT}
\widetilde{w}_k(q) &=& \widetilde{w}(q) \; \; \mathrm{for} \; \; q>k \; , \nonumber \\
\widetilde{w}_k(q) &=& 0 \; \; \mathrm{for} \; \; q<k \; ,
\end{eqnarray}
which, in our theory, corresponds to the \textit{ultra-sharp cut-off}
\begin{eqnarray}
\label{RkHRT}
\widetilde{R}_k(q) &=& 0 \; \; \mathrm{for} \; \; q>k \; , \nonumber \\
\widetilde{R}_k(q) &=& \infty \; \; \mathrm{for} \; \; q<k \; .
\end{eqnarray}
It is the place to mention a relatively unknown paper of Parola\cite{Parola} where the author proposes his own smooth cut-off formulation of HRT theory. Unfortunately his expression for $w_k$ does not belong to the class of pair potentials defined at equation\ (\ref{wk}) so comparisons are hardly possible.
Of course in the framework of the genuine HRT some care is  needed to obtain the flow $\partial_k f_k$ from equation\ (\ref{flow-f}) in the singular ultra sharp cut-off limit. In order to achieve this aim let us rewrite
\begin{eqnarray}
\widetilde{w}_k(q)&=& \Theta_{\epsilon}(q,k)\widetilde{w}(q) \; , \nonumber \\
\partial_k \widetilde{w}_k(q)&=&-\delta_{\epsilon}(q,k)\widetilde{w}(q) \; ,
\end{eqnarray}
where  $\Theta_{\epsilon}(q,k)$ and $\delta_{\epsilon}(q,k)= - \partial_k
\Theta_{\epsilon}(q,k)$ are smoothened versions of the step and Dirac distributions respectively. HRT corresponds to the limit $\epsilon \to 0$ in which $\Theta_{\epsilon}(q,k)\to \Theta(q-k)$ and $\delta_{\epsilon}(q,k)\to
\delta(q-k)$ .
In the case of a homogeneous system the flow of the specific GC free energy is thus given by
\begin{eqnarray}
\label{flow-fHRT}
\partial_{k}f_k(\rho) &=&
 -\frac{1}{2} \int_{q} \frac{ \widetilde{w}(q)\delta_{\epsilon}(q,k) 
}{\widetilde{C}_{k}^{(2)}(q) + \Theta_{\epsilon}(q,k)\widetilde{w}(q) - \widetilde{w}(q)} \; .
\end{eqnarray}
To extract the limit $\epsilon \to 0$ of equation\ (\ref{flow-fHRT}) we will make use of the "little lemma" of Morris\cite{Morris} which states that, for $\epsilon \to 0$ 
\begin{equation}
\label{lemme}
\delta_{\epsilon}(q,k)f( \Theta_{\epsilon}(q,k),k) \to
\delta(q-k) \; \int_{0}^{1}dt \; f(t,q) \;,
\end{equation}
provided that the function $f( \Theta_{\epsilon}(q,k),k)$ is continuous at $k=q$ in the limit $\epsilon \to 0$, which is the case here.
Applying lemma\ (\ref{lemme}) to equation\ (\ref{flow-fHRT}) one gets
\begin{eqnarray}
\label{ij}
\partial_{k}f_k(\rho) &=&  -\frac{1}{2} \int_{q} \widetilde{w}(q)
\delta(q-k) \int_{0}^{1}dt \; \frac{1}{\widetilde{C}_{k}^{(2)} (q)+ (t-1)\widetilde{w}(q)} \nonumber \; , \\
&=&\frac{1}{2} k^{D-1} \frac{S_{D}}{(2 \pi)^{D}}
\ln \left[1 + \widetilde{\mathcal{F}}_k(k) \; \widetilde{w}_{k}  \right] \; ,
\end{eqnarray}
where $S_{D}=2\pi^{D/2}/\Gamma(D/2)$ is the surface of the D-dimensional sphere of unit radius and we have adopted the notations of Parola and Reatto \cite{HRT} $\widetilde{\mathcal{F}}_k(q)=-1/\widetilde{C}_{k}^{(2)}(q)$. Equation\ (\ref{ij}) is nothing but HRT equation for the GC free energy as expected.

Similarly, for the first equation of the hierarchy in k-space, i.e. equation\ (\ref{flow-C1k})
\begin{equation}
\partial_k \widetilde{C}_k^{(1)}(0)= -\lim_{\epsilon \to 0}
\frac{1}{2} \int_q \widetilde{C}_k^{(3)}(q,-q,0)
\frac{\delta_{\epsilon}(q,k)\widetilde{w}(q)}{\left[ \widetilde{C}_{k}^{(2)} (q) + \Theta_{\epsilon}(q,k)\widetilde{w}(q) - \widetilde{w}(q)\right]^{2}} 
\end{equation}
the application of lemma\ (\ref{lemme}) yields a well-defined limit
\begin{equation}
\partial_k \widetilde{C}_k^{(1)}(0)=-\frac{1}{2} k^{D-1} \frac{S_{D}}{(2 \pi)^{D}}\widetilde{C}_k^{(3)}(k,-k,0) 
\frac{\widetilde{\mathcal{F}}_k(k)^{2}\widetilde{w}(k)}{1 + \widetilde{\mathcal{F}}_k(k)\widetilde{w}(k)} \; ,
\end{equation}
which again coincides with the HRT result.

The case $n=2$ is more problematic (and we'll stop here our investigations) since there are two contributions I and II to  the flow  of $\partial_k \widetilde{C}_k^{(2)}(p)=\mathrm{I} + \mathrm{II} $ (cf equation\ (\ref{flow-C2k})). The $\epsilon \to 0$ limit of the first contribution
\begin{equation}
\mathrm{I}= -\frac{1}{2} \int_q \widetilde{C}_k^{(4)}(p,-p,q,-q)\frac{\delta_{\epsilon}(q,k)\widetilde{w}(q)}{\left[ \widetilde{C}_{k}^{(2)} (q) + \Theta_{\epsilon}(q,k)\widetilde{w}(q) - \widetilde{w}(q)\right]^{2}} 
\end{equation}
causes no trouble and application of Morris lemma yields the limit
\begin{equation}
\label{first}
\mathrm{I} =-\frac{1}{2} k^{D-1} \int \frac{d\Omega_{\hat{k}}}{(2 \pi)^{D}} \; \widetilde{C}_k^{(4)}(p,-p,k,-k) \frac{\widetilde{\mathcal{F}}_k(k)^{2}\widetilde{w}(k)}{1 + \widetilde{\mathcal{F}}_k(k)\widetilde{w}(k)} \; ,
\end{equation}
where the angular measure $d\Omega_{\hat{k}}$ denotes the integration over the Euler angles of unit vector $\hat{k}=\vec{k}/k$.
However the second contribution to $\partial_k \widetilde{C}_k^{(2)}(p)$ which may be rewritten as
\begin{eqnarray}
\label{lo}\mathrm{II} &=&\int_q \widetilde{C}_k^{(3)}(p,q,-p-q) \widetilde{C}_k^{(3)}(-p,-q,p+q)
\frac{\delta_{\epsilon}(q,k)\widetilde{w}(q)}{\left[ \widetilde{C}_{k}^{(2)} (q) + \Theta_{\epsilon}(q,k)\widetilde{w}(q) - \widetilde{w}(q)\right]^{2}} \nonumber \times \\
&\times& \frac{1}{\left[ \widetilde{C}_{k}^{(2)} (p+q) + \Theta_{\epsilon}(p+q,k)\widetilde{w}(p+q) - \widetilde{w}(p+q)\right]} 
\end{eqnarray}
causes some troubles. Admittedly the first fraction in the RHS of the above equation is OK and its limit $\epsilon \to0 $ may be taken safely by applying  Morris Lemma but the second one shows a discontinuity for $\parallel \vec{p} + \vec{q} \Vert=k$. However, for a generic $k\neq 0$, the discontinuity occurs only in a zero measure subset of the integration domain. Except exactly at $k=0$ where the integral is not defined, this difficulty would not have any serious consequences as claimed by the authors of reference\cite{HRT}. Following them we define the (discontinuous) function $\widetilde{F}_k(q) =-1/\widetilde{C}_k^{(2)\rightleftarrows}(q)$ for $k>q$ and  $\widetilde{F}_k(q) =-1/(\widetilde{C}_k^{(2)}(q)-\widetilde{w}(q))$ for $k<q$ which allows us to rewrite the second contribution to \ (\ref{lo}) as 
\begin{equation}
\label{second}
\mathrm{II}= -k^{D-1}\int  \frac{d\Omega_{\hat{k}}}{(2 \pi)^{D}}\;
\widetilde{C}_k^{(3)}(p,k,-p-k) \widetilde{C}_k^{(3)}(-p,-k,p+k)\widetilde{F}_k(k+p)\frac{\widetilde{\mathcal{F}}_k(k)^{2}\widetilde{w}(k)}{1 + \widetilde{\mathcal{F}}_k(k)\widetilde{w}(k)} \; .
\end{equation}
where we made use of lemma\ (\ref{lemme}). Collecting the intermediate results\ (\ref{first}) and\ (\ref{second}) we obtain finally
\begin{eqnarray}
\label{total}
-\partial_k \widetilde{C}_k^{(2)}(p)&=&k^{D-1}\int  \frac{d\Omega_{\hat{k}}}{(2 \pi)^{D}}\;\widetilde{C}_k^{(3)}(p,k,-p-k)^{2} \widetilde{F}_k(k+p)\frac{\widetilde{\mathcal{F}}_k(k)^{2}\widetilde{w}(k)}{1 + \widetilde{\mathcal{F}}_k(k)\widetilde{w}(k)}
\nonumber \\
&+&\frac{1}{2} k^{D-1} \int \frac{d\Omega_{\hat{k}}}{(2 \pi)^{D}} \; \widetilde{C}_k^{(4)}(p,-p,k,-k) \frac{\widetilde{\mathcal{F}}_k(k)^{2}\widetilde{w}(k)}{1 + \widetilde{\mathcal{F}}_k(k)\widetilde{w}(k)} \; ,
\end{eqnarray}
in agreement with Parola and Reatto's result. Obviously,  flow equations for the higher order $\widetilde{C}^{(n)}_k$ ($n \geq 3$) will also be ill-defined in the scaling regime $k\to 0$. This observation raises the problem of the validity of HRT in this regime and could be related with the singularities of high ($\sim$ infinite) compressibility states of HRT discussed recently by Reiner\cite{Reiner3}.
\section{\label{RG-KSSHE}The RG in  KSSHE field theory }
These past few years two rigorous scalar field theories have been developed to describe simple fluids:
the collective variable (CV) method introduced by the Ukrainian school\cite{Zuba,Yuk1,Yuk2} and the more recent  KSSHE field theory \cite{Brilliantov,Song,Cai-Mol}.
The CV theory involves two scalar fields with simple physical interpretations while  only one field is involved in KSSHE theory with however a quite obscure physical signification. 
Both theory were shown to be equivalent order by order in the loop-expansion recently\cite{Cai-Pat1,Cai-Pat2}.
In order to relate the RG equations of \mbox{Section II} with RG equations introduced these last past years in  field theory and notably in Wetterich works \cite{Wetterich} the KSSHE framework is at first sight more handy and will be adopted here.
In this framework one can rewrites the GCPF\ (\ref{csi}) of a fluid as a functional integral thanks to a Hubbard-Stratonovich\cite{Hubbard1,Strato} transformation \cite{Cai-Mol}. For attractive pair potentials one has \textit{exactly} (i.e. without unspecified normalization constant)
\begin{eqnarray}
\label{attractive}
\Xi\left[ \nu \right] &=&
\mathcal{N}_{w}^{-1} \int \mathcal{D} \varphi \;
\exp \left( -\frac{1}{2}
\left\langle \varphi \vert w^{-1} \vert \varphi \right\rangle \right)
\Xi_{\text{HS}}\left[ \overline{\nu} + \varphi\right]
 \nonumber \\
 &\equiv &
 \left\langle \Xi_{\text{HS}}\left[ \overline{\nu} + \varphi\right]
 \right\rangle_{w} \; ,
\end{eqnarray}
where $\varphi$ is a real random scalar field, $w^{-1}$ is the inverse of $w$ (in the sense of operators, i.e. $w(1,3) \cdot w^{-1}(3,2)=\delta(1,2)$), and
$\Xi_{\mathrm{HS}}\left[ \overline{\nu} + \varphi\right] $ denotes the GCPF of bare hard
spheres in the presence of a local chemical potential $\overline{\nu}(x) + \varphi(x)$. $\Xi$ can thus be written as a Gaussian average
$\left\langle \ldots \right\rangle_{w}$ of covariance $w$ and we
have noted by $\mathcal{N}_{w}$ the normalization constant
\begin{equation}
 \label{normaw}
 \mathcal{N}_{w}= \int \mathcal{D} \varphi \;  \exp \left( -\frac{1}{2}
\left\langle \varphi \vert w^{-1} \vert \varphi \right\rangle \right) \; .
\end{equation}
The functional integrals which enter
equations\ (\ref{attractive}) and\ (\ref{normaw}) can
be given a precise meaning in the case where the domain $\Omega$
is a cube of side $L$ with periodic boundary conditions, in this case the measure $\mathcal{D} \varphi$ reads as\cite{Wegner}
\begin{subequations}
\label{dphi}
\begin{eqnarray}
\mathcal{D} \varphi & \equiv & \prod_{q \in \Lambda}
\frac{d \widetilde{\varphi}_{q} }
{\sqrt{2 \pi  V}} \\
d \widetilde{\varphi}_{q} d
\widetilde{\varphi}_{-q} & = & 2 \;
d\Re{\widetilde{\varphi}_{q}} \;
d\Im{\widetilde{\varphi}_{q}} \text{ for } q \ne
0 \; .
\end{eqnarray}
\end{subequations}
where $\Lambda = (2 \pi/L)\;  \Z^D$ ($\Z$ ring of integers) is the reciprocal cubic
lattice. Note that the reality of $\varphi$ implies that, for $q\ne
0$, $\widetilde{\varphi}_{-q} =
\widetilde{\varphi}_{q}^{\star}$, where the star means
complex conjugation. With this slick normalization of the functional measure one has exactly
\begin{equation}
\label{NNN}
\mathcal{N}_{w}=\exp\left( \frac{V}{2} \int_{q} \ln \widetilde{w}(q)\right) 
\end{equation}
in the limit of large systems ($L\rightarrow \infty$). Unless some UV cut-off $k_{max}$ is introduced the normalization constant $\mathcal{N}_{w}$ is infinite, but even in the limit $k_{max} \to \infty$ the ratio which defines the GCPF in equation\ (\ref{attractive}) will remain finite and independent of the cut-off for sufficiently large $k_{max}$.

It is instructive to rewrite equation\ (\ref{attractive}) as
\begin{eqnarray}
\label{H-KSSHE}
\Xi\left[ \nu \right] &=&
\mathcal{N}_{w}^{-1} \int \mathcal{D} \varphi \;
\exp \left( -\mathcal{H}_{\mathrm{K}}\left[\nu,\varphi \right] \right) 
 \nonumber \; , \\ 
 \mathcal{H}_{\mathrm{K}}\left[\nu,\varphi \right]&\triangleq&
 \frac{1}{2}
\left\langle \varphi \vert w^{-1} \vert \varphi \right\rangle - \ln \Xi_{\mathrm{HS}}\left[ \overline{\nu} + \varphi \right] \; ,
\end{eqnarray}
where $\mathcal{H}_{\mathrm{K}}\left[\nu,\varphi \right]$ is the tree-level action of the KSSHE field theory since it follows from our dicussions of \mbox{Section II} that at scale-k of the RG group we can rewrite
$\Xi_{k}\left[ \nu \right]$ as
 
\begin{equation}
\label{Start-Point}
\Xi_{k}\left[ \nu \right] =
 \mathcal{N}_{w_{k}}^{-1} \int \mathcal{D} \varphi \;
\exp \left( -\mathcal{H}_{\mathrm{K}}\left[\nu,\varphi \right] 
- \frac{1}{2} \left\langle  \varphi \vert R_{k}\vert \varphi \right\rangle 
\right) 
\; .
\end{equation}
Indeed $w_{k}^{-1}(1,2)= w^{-1}(1,2) + R_{k}(1,2)$ as can be inferred from the definition\ (\ref{wk}) of the pair interaction $w_k$ at scale-k. The presence of the normalization constant  $\mathcal{N}_{w_{k}}$ in the above equation is mandatory in order to obtain the exact flow of $\ln \Xi_k[\nu]$ and $\beta \mathcal{A}_k\left[ \rho \right]$, otherwise they are defined only up to an additional  constant (i.e. which depends only on scale-k and not on the density $\rho$).

Connoisseurs of NPRG theory  will have noticed that the family of models introduced at Equation\ (\ref{Start-Point}) is very similar to that considered by Wetterich\cite{Wetterich,Delamotte} for  generic scalar field theory. His idea was to build a one-parameter  family of models for which a momentum-dependant mass term $\frac{1}{2} \int_q  \widetilde{R}_{k}(q) \widetilde{\varphi}_{q} \;
\widetilde{\varphi}_{-q}$ has been added to the original Hamiltonian (here 
$\mathcal{H}_{\mathrm{K}}\left[\nu,\varphi \right]$) in order to decouple the slow modes by giving them a large mass \cite{Wetterich}. Thus equation\ (\ref{Start-Point}) could have been our starting point to obtain the RG flow of $\ln \Xi_k$ and $\beta \mathcal{A}_k$.
By passing, the flow equation\ (\ref{flow-csik}) for $\ln \Xi_k$ can be obtained in the framework of KSSHE field theory without any recourse to the GC formalism. This somehow redundant material is left as an exercise for the reader.

A more detailed comparison of the NPRG  of section\ II with the NPRG for generic (or canonical)  field theory is worthwile but not so trivial. We have pointed out the similarities of the two RG constructions but there is a fundamental difference. The KSSHE theory is not a canonical scalar field theory in the sense that the coupling between the external source $\nu(x)$ and the field $\varphi(x)$ is non-linear as apparent in equation\ (\ref{H-KSSHE}). In order to establish the exact link between the RG group of Section\ II and the NPRG of Wetterich we need first to build a canonical field theory for simple fluids, this is the aim of next section.
\section{\label{RG-canonic}The RG in a canonical field theory of simple fluids }
\subsection{\label{canonic} A canonical field theory for simple fluids}
The starting point will be the equation\ (\ref{H-KSSHE}) for $\Xi\left[\nu \right]$ that we rewrite with a slight change of notations ($\chi$ instead of $\varphi$ to avoid further confusions) as

\begin{equation}
 \Xi\left[ \nu \right] =
\mathcal{N}_{w}^{-1} \int \mathcal{D} \chi \;
\exp \left( -\frac{1}{2}
\left\langle \chi \vert w^{-1} \vert \chi \right\rangle
 + \ln \Xi_{\text{HS}}\left[ \overline{\nu} + \chi \right]
 \right)
\end{equation}
Let us make now the change of variables\cite{Cai-Mol}
\begin{equation}
\overline{\nu} + \chi = \nu_{0} + \varphi
\end{equation}
where $\nu_{0}$ is an arbitrary constant, hence $\varphi=\chi + \Delta \nu$ with $\Delta \nu=\nu+\nu_{\mathrm{S}}-\nu_{0}$. With this trick one is led to a canonical field theory defined by its $\nu$-independent Hamiltonian
\begin{equation}
\label{Hcano}
\mathcal{H}\left[\varphi \right] =\frac{1}{2}\left\langle \varphi \vert w^{-1}\vert
\varphi \right\rangle -\ln \Xi_{\text{HS}}\left[ \nu_{0} + \varphi \right] \; ,
\end{equation}
and its partition function
\begin{equation}
\Xi^{\star}\left[ J\right]=\mathcal{N}_{w}^{-1} \int \mathcal{D} \varphi \;
\exp \left( - \mathcal{H}\left[\varphi \right] + \left\langle J \vert \varphi
\right\rangle \right) 
 \; ,
\end{equation}
where the external source 
\begin{equation}
\label{Jnu}
J(1)=w^{-1}(1,2) \cdot \Delta \nu(2) \; .
\end{equation}
Denoting by $W^{\star}\left[ J \right]= \ln \Xi^{\star}\left[ J\right]$ and
$W\left[ \nu \right]= \ln \Xi \left[ \nu\right]$ the grand potential (up to a multiplicative $ -k_{B}T$) we have of course
\begin{subequations}
\label{WWstar}\begin{eqnarray}
W^{\star}\left[ J \right]&=& W\left[ \nu \right] +\dfrac{1}{2} \left\langle \Delta\nu  \vert w^{-1} \vert \Delta\nu \right\rangle\; \forall J , \; \;  \nu \; \mathrm{given \; by} \; \Delta \nu=w\cdot J \; , \\
W\left[ \nu \right]&=&W^{\star}\left[ J \right] -  \dfrac{1}{2}\left\langle J \vert w \vert J \right\rangle  \; \forall \nu , \; \; J \; \mathrm{given \; by} \; J=w^{-1}\cdot \Delta \nu \; .
\end{eqnarray}
\end{subequations}
The above relations may be used to establish the relations between the density correlation functions $G^{(n) \; T}$ and the truncated correlation functions of the field $\varphi$, $G_{\varphi}^{(n) \; T}(1,\ldots,n)=\delta W^{\star}\left[J \right]/\delta J(1) \ldots \delta J(n) $. Noting that it follows from equation\ (\ref{Jnu}) that  $\delta \left[ \ldots \right] / \delta J(1) =
w(1,2)\cdot\delta\left[  \ldots \right] / \delta \nu(2)$ on obtains readily from equations\ (\ref{WWstar}) that
\begin{subequations}
\label{toto}\begin{eqnarray}
\label{toto1}\Phi(1)&=&\left\langle \varphi(1)\right\rangle =w(1,2)\cdot \rho(2) + \Delta \nu(1) \; , \\
G_{\varphi}^{(2) \; T}(1,2)&=&w(1,1^{'}) w(2,2^{'}) \;
G^{(2) \; T}(1^{'},2^{'}) \; \; + \;w(1,2) \\
G_{\varphi}^{(n) \; T}(1,\ldots,n)&=&w(1,1^{'}) \ldots w(n,n^{'}) \;
G^{(n) \; T}(1^{'},\ldots,n^{'}) \; \;\forall n \geq 3 \; .
\end{eqnarray}
\end{subequations}
Since the KSSHE field $\chi$ and $\varphi$ differ by a constant their truncated correlation functions coincide for $n\geq 2$, and, for $n=1$ of course $\Phi(1)=\left\langle \chi(1) \right\rangle_{\mathrm{K}} +\Delta \nu(1)$ from which the relation $\rho(1)=w^{-1}(1,2)\left\langle \chi(2) \right\rangle_{\mathrm{K}}$ follows (the subscript $\left[ \ldots\right]_{\mathrm{K}}$ means that the statistical average must be computed in the framework of KSSHE field theory). Therefore equations\ (\ref{toto})  establish the link between the GC formalism and the KSSHE field theory as well.

We turn now our attention to the Gibbs free energy $\Gamma\left[ \Phi \right]$ which will be  defined as usual as   the Legendre transform of the convex functional $W^{\star}\left[ J\right]$\cite{Zinn} :   
\begin{eqnarray}
\label{Gamma}
(\forall \phi \in \mathcal{F}) \; \;\Gamma\left[ \Phi\right] &=& \sup_{J\in \mathcal{J}}\left\lbrace  \left\langle J \vert\Phi\right\rangle -W^{\star}\left[ J\right]\right\rbrace \;, \nonumber \\
&=& \left\langle J^{\star} \vert\Phi\right\rangle -W^{\star}\left[ J^{\star}\right]  
\end{eqnarray}
where $J^{\star}$ is the unique solution of the implicit equation
\begin{equation}
\left. \frac{\delta W^{\star}\left[ J\right] }{\delta J(1)}\right\vert_{J=J^*}=\Phi(1) \; ,
\end{equation}
and $\mathcal{F}$ and $\mathcal{J}$ are the convex sets of fields and  sources respectively. One has then of course $J(1)=\delta\Gamma\left[ \Phi\right]/ \delta \Phi(1)$. 

We will now establish the link between the functionals $\Phi \longmapsto \Gamma\left[ \Phi\right]$ and $\rho \longmapsto
\beta \mathcal{A}\left[ \rho \right]$. To unclutter the notations we will consider rather
\begin{equation}
\beta \overline{\mathcal{A}}\left[ \rho \right] \triangleq \beta
\mathcal{A}\left[ \rho \right] \; + \; \left\langle \nu_{\mathrm{S}}- \nu_{0}\vert \rho \right\rangle \; ,
\end{equation}
so that $\beta \overline{\mathcal{A}}\left[ \rho \right]=\sup_{\rho \in \mathcal{R}}\left\lbrace \left\langle \rho \vert \Delta \nu \right\rangle -W\left[ \nu\right]  \right\rbrace $ and $\Delta \nu(1)=\delta \beta \overline{\mathcal{A}}\left[ \rho \right] /\delta \rho(1)$ (since $\beta \overline{\mathcal{A}}\left[ \rho \right]$ and $\beta
\mathcal{A}\left[ \rho \right]$ differ by a linear functional, $\beta \overline{\mathcal{A}}\left[ \rho \right]$  is also a convex functional of the local density). In the one hand the relation\ (\ref{toto1}) between the order parameters $\rho$ and $\Phi$ can be rewritten as
\begin{eqnarray}
\Phi(1)&=& w(1,2)\cdot \rho(2) \;  + \; \frac{\delta  \overline{\mathcal{A}}\left[\rho \right] }{\delta \rho(1)} \nonumber \; , \\
\rho(1)&=& w^{-1}(1,2)\cdot \Phi(2) \; - \; \frac{\delta \Gamma\left[\Phi \right] }{\delta \Phi(1)} \; ,
\end{eqnarray}
and, on the other hand, replacing $W^{\star}\left[J \right]$ by its expression\ (\ref{WWstar}) in term of the grand potential $W\left[\nu \right]$ yields the useful identity
\begin{equation}
\label{identite}
\beta \overline{\mathcal{A}}\left[ \rho \right]=\Gamma\left[\Phi \right] \;
-\frac{1}{2}  \left\langle J \vert \Delta \nu\right\rangle ,
\end{equation}
from which we deduce that either
\begin{eqnarray}
\label{AGam}(\forall \rho \in \mathcal{R}) \; \;\beta \overline{\mathcal{A}}\left[ \rho \right]&=&\Gamma\left[\Phi \right] -\frac{1}{2}  
\left\langle  \frac{\delta \Gamma\left[ \Phi\right] }{\delta \Phi}\vert w  \vert\frac{\delta \Gamma\left[ \Phi\right] }{\delta \Phi}\right\rangle \;, \nonumber \\
\Phi \;  \mathrm{sol.} \;  \rho(1)&=& w^{-1}(1,2) \cdot \Phi(2) -\frac{\delta \Gamma\left[ \Phi\right] }{\delta \Phi(1)} \; ,
\end{eqnarray}
where "sol." means "solution of", or 
\begin{eqnarray}
\label{GamA}(\forall \Phi \in \mathcal{F}) \; \; \Gamma\left[\Phi \right]&=&\beta \overline{\mathcal{A}}\left[ \rho \right] +\frac{1}{2}  
\left\langle  \frac{\delta \beta \overline{\mathcal{A}}\left[ \rho \right] }{\delta \rho}\vert w^{-1}  \vert\frac{\delta \beta \overline{ \mathcal{A}}\left[ \rho \right] }{\delta \rho}\right\rangle \;, \nonumber \\
\rho \;  \mathrm{sol.} \;  \Phi(1)&=& w(1,2) \cdot \rho(2) + \frac{\delta \beta \overline{ \mathcal{A}}\left[ \rho \right] }{\delta \rho(1)}  \; .
\end{eqnarray}

Finally the vertex functions of the canonical theory will be defined as usual \cite{Zinn}
\begin{equation}
\label{vertex}
\Gamma^{(n)}\left[\Phi \right](1,\ldots,n) = \frac{\delta \Gamma\left[\phi \right] }{\delta \Phi(1) \ldots \delta\Phi(n)} \; . 
\end{equation}

It is also possible to establish the general relations between the $\Gamma^{(n)}\left[\Phi \right](1,\ldots,n)$ and the direct correlation functions of the fluid $C^{(n)}\left[\rho \right](1,\ldots,n)$ but they look rather awkward. By the way, only the cases $n=1,2$ will be useful, we quote them without the proofs which are elementary
\begin{eqnarray}
\label{skonz} C^{(1)}(1)&=&w^{-1}(1,2) \cdot \Gamma^{(1)}(2)  \; , \nonumber \\
 C^{(2)}(1,3)\cdot\Gamma^{(2)}(3,2)&=&
  w^{-1}(1,3)\cdot C^{(2)}(3,2) +
 w(1,3)\cdot \Gamma^{(2)}(3,2)  
  \; .  
\end{eqnarray}

\subsection{\label{canonicRG} The RG flow in the canonical theory}
 Following Berges \textit{et al.}\cite{Wetterich} one defines
\begin{equation}
 \Xi_{k}^{\star}\left[ J \right] =
 \mathcal{N}_{w_{k}}^{-1} \int
 \mathcal{D}\varphi \; \exp\left( -\mathcal{H}\left[\varphi \right] -
 \frac{1}{2} \left\langle  \varphi \vert R_{k} \vert \varphi \right\rangle
 +\left\langle  J \vert \varphi \right\rangle \right) \; ,
\end{equation}
where $\mathcal{H}\left[\varphi \right]$ is the Hamiltonian\ (\ref{Hcano}) of the canonical theory of the fluid, $R_k$ is the regulator defined in Section \ref{basics}, and $\mathcal{N}_{w_{k}}$ the normalization constant\ (\ref{normaw}) (with the replacement $w \rightarrow w_k$). The normalization constant $\mathcal{N}_{w_{k}}$ is rarely considered in the literature on statistical field theory but here, it ensures that $\Xi_{k}^{\star}\left[ J \right]$ is \textit{exactly} the partition function of the canonical field theory of a HS fluid with $w_k$ pair interactions the bare action of which is given by   $\mathcal{H}_{k}\left[\varphi \right]=\mathcal{H}\left[\varphi \right]+ \left\langle  \varphi \vert R_k \vert \varphi \right\rangle /2$.
 
The effective average action $\Gamma_k\left[ \Phi \right]$ is defined as\cite{Wetterich0,Wetterich}
\begin{equation}
 \Gamma_{k}^{'}\left[\Phi \right]=
 \sup_{J\in\mathcal{J}}\left\lbrace  \left\langle J \vert
 \Phi \right\rbrace -W^{\star}_{k}\left[ J\right]\right\rbrace  \; ,   
\end{equation}
(where $W^{\star}_{k}\left[ J\right]=\ln \Xi_{k}^{\star}$) by the relation
\begin{equation}
\label{ppa}
\Gamma_{k}^{'}\left[\Phi \right]=\Gamma_{k}\left[\Phi \right]+
\frac{1}{2} \left\langle \Phi \vert R_k \vert \Phi \right\rangle \; .
\end{equation}
Note that $\Gamma_{k}^{'}\left[\Phi \right]$ is convex while $\Gamma_{k}\left[\Phi \right]$ is not.
 
In the limit $k \to 0$, $R_k \to 0$ and thus $\Gamma_{k=0}^{'}=\Gamma_{k=0}= \Gamma$.
In order to obtain the $k\to\infty$ limit of $\Gamma_{k}$ we start from its functional implicit expression\cite{Wetterich} 
\begin{eqnarray}
\label{limitinfty}
\exp\left( -\Gamma_{k}\left[ \Phi \right]  \right)&=& 
\mathcal{N}_{w_{k}}^{-1} \int \mathcal{D} \chi \; 
\exp\left( \mathcal{H}\left[ \Phi + \chi\right] +
\left\langle  \chi \vert \frac{\delta \Gamma_{k}\left[\Phi \right]}{\delta \Phi }\right\rangle  -\frac{1}{2} \left\langle \chi \vert
R_k \vert \chi \right\rangle   \right) \;.
\end{eqnarray}
For $k \to \infty$, $R_k \to \infty$ and therefore $\widetilde{w}_k\sim R_{k}^{-1}$. The point is that  we have exactly, in the limit $R_k \to \infty$, without unspecified multiplicative constant, 
\begin{equation}
\label{deltaf}
\frac{1}{\mathcal{N}_{R_{k}^{-1}}}\exp\left( 
-\frac{1}{2}
\left\langle \chi \vert R_k \vert \chi\right\rangle \right) =\delta_{\mathcal{F}}\left[\chi \right]  \; ,
\end{equation}
where the functional Dirac mass $\delta_{\mathcal{F}}\left[\chi \right]$ is given, with Wegner's normalization\cite{Wegner,Cai-Pat1,Cai-Pat2} by
\begin{equation}
\label{deltaF} \delta_{\mathcal{F}}\left[ \chi \right]=
\sqrt{2 \pi V} \; \delta \left( \widetilde{\chi }_0 \right)  \;
\prod_{q \in \Lambda \; , q_x>0}\left[ \pi V \; \delta
\left( \Re{\widetilde{\chi}_{q}} \right)  \; \delta
\left( \Im{\widetilde{\chi}_{q}} \right) \right] \; ,
\end{equation}
as a direct calculation will show. Reporting equation\ (\ref{deltaf}) in\ (\ref{limitinfty}) we obtain $\Gamma_{k=\infty}\left[\Phi \right] = \mathcal{H}\left[ \Phi \right]$.

The RG flow of $\Gamma_{k}\left[ \Phi \right]$ is given by the equation\cite{Wetterich}
\begin{eqnarray}
\label{flowGamma}
\partial_{k} \Gamma_{k}\left[ \Phi \right]& =&
\frac{1}{2} \int d1\; d2 \;\partial_{k} R_{k}(1,2) 
\left\lbrace R_k + \Gamma_{k}^{(2)}\right\rbrace^{-1}(1,2) \nonumber \\
& - & \frac{1}{2} \int d1\; d2 \; \partial_{k} R_{k}(1,2)w_k(1,2) \; ,
\end{eqnarray}
where $\Gamma_{k}^{(2)}$ is one of the effective vertex functions
\begin{equation}
\Gamma_{k}^{(n)}\left[ \Phi\right](1,\ldots,n)=
\frac{\delta \Gamma_{k}\left[ \Phi \right]}{\delta \Phi(1) \ldots \delta \Phi(n)} \; . 
\end{equation}
The first contribution  to  the right hand side of equation\ (\ref{flowGamma}) seems to have been obtained for the first time by Nicoll \cite{Nicoll} and rediscovered later by Wetterich\cite{Wetterich0} while the second one is not considered in general and stems from the trivial flow of the normalization constant $\mathcal{N}_{w_k}$. Indeed it follows from equation\ (\ref{NNN}) that
\begin{eqnarray}
\partial_k \ln \mathcal{N}_{w_k} &=& \frac{V}{2} \; \partial_k \int_q \ln\widetilde{w}_k(q) \nonumber \\
&=& -\frac{V}{2} \int_q \widetilde{R}_k(q)\widetilde{w}_k(q) \nonumber \\
&=&
-\frac{1}{2}\int d1 d2 \;\partial_k R_k(1,2) \; w_k(1,2) \; . 
\end{eqnarray}

The equations for the flows  $\partial_k\Gamma_{k}^{(n)}\left[ \Phi\right](1,\ldots,n)$ of  the running vertex are  obtained by taking the functional derivatives of equation\ (\ref{flowGamma}) with respect to $\Phi(x)$\cite{Wetterich}. They are obviously identical to those derived in section\  (\ref{flow-direct}) for the effective correlation functions under the replacements $C_k^{(n)} \longrightarrow - \Gamma_{k}^{(n)}$ and $\partial_k w_k \longrightarrow \partial_k R_k$ for the vertex  and $G^{(2),\; T}=-\left\lbrace C^{(2)}_k +w_k -w \right\rbrace^{-1} \longrightarrow
G^{(2),\; T}_{\varphi}=\left\lbrace \Gamma_{k}^{(2)} + R_k\right\rbrace^{-1}$ for the propagators.

Until now we have only considered the first step of Wilson program\cite{Wilson}, i.e. the coarse graining operation. In order to get fixed points and apply the whole machinery of the RG group theory to the study of critical phenomena we have to consider now the second step, i.e. a rescaling of distances, wave vectors and fields. We refer the reader to the abundant literature on this point\cite{Wilson,Wetterich,Bervillier,HRT,Delamotte}, and only quote the RG equation obtained after rescaling
\begin{equation}
 \label{fullRG}\partial_t \Gamma_t =\mathcal{G}_{\mathrm{dil}}\Gamma_t +
  \frac{1}{2} \int d1\; d2 \;\partial_{t} R_{t}(1,2) 
\left\lbrace R_t + \Gamma_{t}^{(2)}\right\rbrace^{-1}(1,2) \; ,
\end{equation}
where $t=\ln(k/k_{\mathrm{max}})$ is the RG "time" and the expression of the dilatation operator may be written as\cite{Bervillier}
\begin{equation}
\mathcal{G}_{\mathrm{dil}} =\int_q \vec{q} \cdot \partial_{\vec{q}}\widetilde{\Phi}_q \;\frac{\delta}{\delta\widetilde{\Phi}_q} + (D-D_{\Phi})\int_q
 \widetilde{\Phi}_q \frac{\delta}{\delta\widetilde{\Phi}_q} \; ,
\end{equation}
where $D_{\Phi}=(D-2 +\eta)/2$ is the anomalous dimension of the field $\Phi$ (recall that Fisher's exponent $\eta$ is a positive constant defined with respect to a non trivial fixed point of the flow\ (\ref{fullRG})). Note that  we have discarded the trivial flow of the renormalization constant $\mathcal{N}_t$ from equation\ (\ref{fullRG}) to display an uncluttered  formula. The flow of $\beta\mathcal{A}_k$ can be treated  in a similar way\cite{HRT}.

We want to point out that by contrast with the flow equations obtained in the  GC or KSSHE theories, that obtained in the canonical theory of liquids is independent of the pair interaction $w(r)$. All simple fluids are thus described by the same flow, they differ only by the initial conditions at $k=\infty$ which prescribe the microscopic model, i.e the pair potential and the reference system (something else than hard spheres can be chosen). This property together with the existence of fixed points and a group law for the flow\ (\ref{fullRG}) yield the universality of critical properties of all simple fluids with short range interactions. Their universality class is of course that of the Ising model.
\subsection{\label{lien} The relation between the flows of $\Gamma_{k}\left[ \Phi \right] $ and $\beta \mathcal{A}_k\left[ \rho \right] $}
It transpires from the previous sections that the flows of  $\Gamma_{k}\left[ \Phi \right] $ and $\beta \mathcal{A}_k\left[ \rho \right] $ must be related in some way since both describe the same physics. We first note that the k-system being a HS fluid with pair interactions $w_k(r)$ the relations established in Section\ (\ref{canonic}) between the "true" action $\Gamma_{k}^{'}$ and the "true" GC free energy $\beta \mathcal{A}_{k}^{'}$  can be applied at scale-k and one has for instance (cf equation\ (\ref{AGam}))

\begin{eqnarray}
\label{ApkGampk}\beta \overline{\mathcal{A}}_{k}^{'}\left[ \rho \right]&=&\Gamma_{k}^{'}\left[\Phi \right] -\frac{1}{2}  
\left\langle  \frac{\delta \Gamma_{k}^{'}\left[ \Phi\right] }{\delta \Phi}\vert w_k  \vert\frac{\delta \Gamma_{k}^{'}\left[ \Phi\right] }{\delta \Phi}\right\rangle \;, \nonumber \\
\Phi \;  \mathrm{sol.} \;  \rho(1)&=& w_{k}^{-1}(1,2) \cdot \Phi(2) -\frac{\delta \Gamma_{k}^{'}\left[ \Phi\right] }{\delta \Phi(1)} \; ,
\end{eqnarray}
where 
\begin{equation}
\beta\overline{\mathcal{A}}_{k}^{'}\left[ \rho \right]\triangleq 
\beta \mathcal{A}_{k}^{'}\left[ \rho \right] + \left\langle \nu_{\mathrm{S},\; k} -\nu_0\vert \rho \right\rangle \; .
\end{equation}
Recall that $\beta \mathcal{A}_{k}^{'}\left[ \rho \right]$ and $\beta \mathcal{A}_{k}\left[ \rho \right]$ are related through equation\ (\ref{def-Ak}). Now we define
\begin{equation}
\beta\label{jambon}\overline{\mathcal{A}}_{k} \left[ \rho \right]\triangleq 
\beta \mathcal{A}_{k} \left[ \rho \right] + \left\langle \nu_{\mathrm{S}} -\nu_0\vert \rho \right\rangle \; 
\end{equation}
so that equation\ (\ref{def-Ak}) can be rewritten as
\begin{equation}
\label{ppo}
\beta\overline{\mathcal{A}}_{k} \left[ \rho \right]=\beta\overline{\mathcal{A}}_{k}^{'}\left[ \rho \right] +
\frac{1}{2} \left\langle \rho \vert w_k -w \vert \rho \right\rangle \;.
\end{equation}
Reexpressing  equations\ (\ref{ApkGampk}) in terms of $\Gamma_k$ and $\overline{\mathcal{A}}_{k}$ thanks to equations\ (\ref{ppa}) and\ (\ref{ppo}) yields
\begin{eqnarray}
\label{AkGamk}\beta \overline{\mathcal{A}}_{k}\left[ \rho \right]&=&\Gamma_{k}\left[\Phi \right] -\frac{1}{2}  
\left\langle  \frac{\delta \Gamma_{k}\left[ \Phi\right] }{\delta \Phi}\vert w  \vert\frac{\delta \Gamma_{k}\left[ \Phi\right] }{\delta \Phi}\right\rangle \;, \nonumber \\
\Phi \;  \mathrm{sol.} \;  \rho(1)&=& w^{-1}(1,2) \cdot \Phi(2) -\frac{\delta \Gamma_k\left[ \Phi\right] }{\delta \Phi(1)} \; .
\end{eqnarray}
This remarkable result shows that the functional relations between the running $\Gamma_k$ and $\beta \overline{\mathcal{A}}_{k}$ as well as those  between the order parameters $\rho$ and $\Phi$ are in some sense independent of scale-k.
Proceeding in the same way with equation\ (\ref{GamA}) one finds the dual relation
\begin{eqnarray}
\label{GamkAk}\Gamma_k\left[\Phi \right]&=&\beta \overline{\mathcal{A}}_k\left[ \rho \right] +\frac{1}{2}  
\left\langle  \frac{\delta \beta \overline{\mathcal{A}}_k\left[ \rho \right] }{\delta \rho}\vert w^{-1}  \vert\frac{\delta \beta \overline{ \mathcal{A}}_k\left[ \rho \right] }{\delta \rho}\right\rangle \;, \nonumber \\
\rho \;  \mathrm{sol.} \;  \Phi(1)&=& w(1,2) \cdot \rho(2) + \frac{\delta \beta \overline{ \mathcal{A}}_k\left[ \rho \right] }{\delta \rho(1)}  \; .
\end{eqnarray}
Note that the mapping $\rho \mapsto \Phi$ is a one to one correspondence since we have, for instance from equation\ (\ref{GamkAk}), that $\delta \Phi(1)/\delta \rho(2)=w(1,2)-C^{(2)}_k(1,2)$, which is a  positive definite operator since it follows from equation\ (\ref{sign-den})  that $\widetilde{w}(q)-\widetilde{C}^{(2)}_k(q)>\widetilde{w}_k(q) >0$. For homogeneous systems the order parameter $\Phi$ is therefore an increasing \textit{function} of the density $\rho$.
As a consequence of equations\ (\ref{AkGamk}) and\ (\ref{GamkAk}) we also have
\begin{equation}
\label{31}
\frac{\delta\beta\overline{\mathcal{A}}_k\left[ \rho \right] }{\delta \rho(1)}=
w(1,2) \cdot \frac{\delta\Gamma_k\left[ \Phi \right] }{\delta \Phi(2)} \; ,
\end{equation}
or equivalently $C_k^{(1)}(1)=-w(1,2) \cdot \Gamma^{(1)}_k(2)$. This relation between $n=1$ vertex functions and direct correlation functions happens once again to be an identity independent of scale-k.

We are now in position to establish the relation between the two flows $\partial_k \beta \overline{\mathcal{A}}_k\left[ \rho \right]$ and
$\partial_k  \Gamma_k\left[ \Phi \right]$. Taking the partial derivative of equation\ (\ref{AkGamk}) with respect to $k$ at \textit{fixed} $\rho$, one finds that
\begin{eqnarray}
\label{uj}
\partial_k \beta \overline{\mathcal{A}}_k\left[ \rho \right]&=&
\left. \partial_k  \Gamma_k\left[ \Phi \right]\right\vert_{\rho} -
\left\langle \frac{\delta \beta \overline{\mathcal{A}}_k}{\delta \rho} \left\vert
w^{-1} \right\vert \partial_k\frac{\delta \beta \overline{\mathcal{A}}_k} {\delta \rho}\right\rangle \; , \nonumber \\
&=& \partial_k  \Gamma_k\left[ \Phi \right]+
\left\langle \frac{\delta \Gamma_k}{\delta \Phi}\vert \partial_k\Phi\vert_{\rho} \right\rangle  -
\left\langle \frac{\delta \beta \overline{\mathcal{A}}_k}{\delta \rho} \left\vert
w^{-1} \right\vert \partial_k\frac{\delta \beta \overline{\mathcal{A}}_k} {\delta \rho}\right\rangle \; , \nonumber \\
&=& \partial_k  \Gamma_k\left[ \Phi \right] +
\left\langle \partial_k \frac{\delta \beta \overline{\mathcal{A}}_k}{\delta \rho} \vert 
w^{-1}\cdot 
\frac{\delta \beta \overline{\mathcal{A}}_k}{\delta \rho}-
\frac{\delta \Gamma_k}{\delta \Phi}\right\rangle  \; ,
\end{eqnarray}
where the last line follows from
\begin{equation}
\partial_k\frac{\delta \beta \overline{\mathcal{A}}_k}{\delta \rho(1)}=
\partial_k\Phi(1)\vert_{\rho} \; ,
\end{equation}
as can be inferred from equation\ (\ref{GamkAk}). Note that the second term in the right hand side of the last line of equation\ (\ref{uj}) vanishes by virtue of equation\ (\ref{31}). Since $\partial_k \beta \mathcal{A}_k\left[ \rho \right]=\partial_k \beta \overline{\mathcal{A}}_k\left[ \rho \right]$ (cf equation\ (\ref{jambon})) we have finally
\begin{eqnarray}
\label{egaflow}
\partial_k \beta\mathcal{A}_k\left[ \rho \right] &=&\partial_k \Gamma_k\left[ \Phi \right] \; , \nonumber \\
(\forall \rho ) , \; \;  \Phi \;  \mathrm{sol.} \;  \rho(1)&=& w^{-1}(1,2) \cdot \Phi(2) -\frac{\delta \Gamma_k\left[ \Phi\right] }{\delta \Phi(1)} \; , \nonumber \\
& \mathrm{ or} & \; \nonumber \\
(\forall \Phi ), \; \; \rho \;  \mathrm{sol.} \;  \Phi(1)&=& w(1,2) \cdot \rho(2) + \frac{\delta \beta  \mathcal{A}_k\left[ \rho \right] }{\delta \rho(1)} +\nu_S -\nu_0 \; .
\end{eqnarray}
Therefore the two flows coincide, moreover it is not difficult to check that the initial conditions $\beta \mathcal{A}_{\infty}\left[ \rho \right]\equiv \beta \mathcal{A}_{\mathrm{MF}}\left[ \rho\right] $ and $\Gamma_{\infty}\left[ \Phi \right]\equiv \mathcal{H} \left[  \Phi \right]$ are of course also related by relations\ (\ref{egaflow}) as a short calculation will show.
\subsection{\label{comments}Additional Comments}
We complete the results of Section\ (\ref{lien}) with a study of correlation functions at scale-k. The relations\ (\ref{toto}) are of course valid for the k-systems, so that we have
\begin{subequations}
\label{totok}\begin{eqnarray}
\label{totok1}G_{\varphi, \; k}^{(2) \; T}(1,2)&=&w_k(1,1^{'}) w_k(2,2^{'}) \;
G^{(2) \; T}_k(1^{'},2^{'}) \; \; + \;w_k(1,2) \\
G_{\varphi, \; k}^{(n) \; T}(1,\ldots,n)&=&w_k(1,1^{'}) \ldots w_k(n,n^{'}) \;
G^{(n) \; T}_k(1^{'},\ldots,n^{'}) \; \;\forall n \geq 3 \; ,
\end{eqnarray}
\end{subequations}
where the $G_{\varphi, \; k}^{(n) \; T}$ are the correlation functions of the KSSHE field at scale-k.
The  relations\ (\ref{totok}) may be used to derive the identity of the flows\ (\ref{egaflow}) starting from equation\ (\ref{flowA}) for $\partial_k \beta \mathcal{A}_k$ by replacing $G_k^{(2), \; T}$ by $G_{\varphi, \; k}^{(2), \; T}$ by its expression\ (\ref{totok1}); one then obtains the expression\ (\ref{flowGamma}) for $\partial_k \Gamma_k$ (including the flow of the normalization $\mathcal{N}_{w_k}$) as the reader will check easily.

Concerning the vertex functions we have already given the relation between $C^{(n=1)}_{k}$ and $\Gamma^{(n=1)}_{k}$ (cf equation\ (\ref{31})). For  $n=2$ and  at scale-k we have from equation\ (\ref{skonz})
\begin{equation}
C_k^{'(2)}(1,3)\cdot\Gamma_k^{'(2)}(3,2)= w_k^{-1}(1,3)\cdot C_k^{'(2)}(3,2) +
 w_k(1,3)\cdot \Gamma_k^{'(2)}(3,2) \; ,
\end{equation}
from which it is easy infer that
\begin{equation}
C_k^{(2)}(1,3)\cdot\Gamma_k^{(2)}(3,2)= w^{-1}(1,3)\cdot C_k^{(2)}(3,2) +
 w(1,3)\cdot \Gamma_k^{(2)}(3,2) \; .
\end{equation}
It is likely -but we did not try to prove it- that the relations between the $C_k^{(n)}$ and $\Gamma_k^{(n)}$ are also independent of the scale-k  at any order '$n$', i.e. they involve $w$ rather than  $w_k$.

From this lengthy discussion on the identity of the flows  $\partial_k\beta \mathcal{A}_k$ and $\partial_k \Gamma_k$ we can draw general conclusions which enlighten some obscure points related to the singular behavior of the HRT theory in the subcritical region. Known properties of the flow of $\Gamma_k$ can be transposed to the flow of $\beta \mathcal{A}_k$  thanks to the exact equations\ (\ref{AkGamk}) and\ (\ref{egaflow}). For instance, it is known from  theoretical and numerical studies of many different models treated in various approximation schemes that, in general, $\partial^2 \Gamma_k/\partial \Phi^2$ is discontinuous at the coexistence curve below $T_c$, jumping from a finite value in the ordered phase to zero in the inner region of the coexistence curve. In other words, the susceptibility jumps from a finite value to infinity. This property  survives the one-to-one mapping\ (\ref{AkGamk}) yielding the conclusion that, for simple fluids, the second derivative of the free energy $\partial^2 f_k / \partial \rho^2$ will also exhibits a finite discontinuity on the binodal, being zero in the two phase region, as a short calculation will show.  By contrast the HRT yields a continuous $\partial^2 f_k / \partial \rho^2 \; \; $ \cite{HRT,HRT2,Reiner1,Reiner2} and thus, in this framework, the binodal and the spinodal coincide. It is likely that the ultra sharp cut-off limit needed to extract HRT from our smooth cut-off equations enforces the continuity of $\partial^2 f_k / \partial \rho^2$. One could argue that this flaw of the theory could also be due to the approximations injected into the exact HRT flow hierarchy in order to solve it. We do not think so and  support the former hypothesis basing our point of view 
i) on the fact that the discontinuity of $\partial^2 \Gamma_k/\partial \Phi^2$ at subcritical temperatures is observed whatever the approximation retained\cite{Wetterich}
ii) on the analysis of Section\ (\ref{flow-direct}) showing that the HRT equations for the two-body and higher-order direct correlation functions are ill-defined in the scaling regime $k\to 0$. If this is true then there is no way to improve HRT even by incorporating sophisticated closure relations such as those discussed in reference\cite{Reiner3}.
Unfortunately, in his own smooth cut-off formulation of HRT (which however cannot  reduce  to our formulation in any way, as discussed in Section\ (II)) Parola does not discuss this point\cite{Parola}. 
We cannot help raising the doubt: is HRT really intrinsically correct?
\section{\label{KAC}RG analysis of Kac model }
As an illustration of the issues of previous sections we study here the RG flow of Kac model which can done analytically. 
\subsection{\label{KAC1} Generalities}
In the Kac model the hard spheres interact via a long range pair potential $w(r)$ obtained as the limit\cite{Hansen,Kac,Kac1,Kac2,Kac3}
\begin{equation}
w(r)= \lim_{\alpha \to 0} \; \alpha^D \Phi(\alpha r) \; \; \mathrm{ at \; fixed \; volume \; V\;} \;,
\end{equation}
where $\Phi(r)$ is a short-range function, distances are measured in unit of $\sigma$ and the inverse range of the interaction $\alpha>0$ is dimensionless. We assume $\widetilde{\Phi}(k)>0$ and $\Phi(0)$ finite and note that $\widetilde{w}(k)=  \widetilde{\Phi}(k/\alpha)$.  It is important to emphasize that the limit $\alpha \to 0$ must be taken \textit{before}  taking the thermodynamic limit, i.e. at fixed finite volume $V$. Assuming PBC conditions the potential  energy $E_p(\mathcal{C})$ of any GC configuration $\mathcal{C}$ will be 
\begin{equation}
-E_p(\mathcal{C})/k_B T=\frac{1}{2}\frac{1}{V}\sum_{k\in\Lambda}\widetilde{w}(k) 
\widetilde{\rho}(k)\widetilde{\rho}(-k) \; ;
\end{equation}
where $\widetilde{\rho}(k)$ is the Fourier transform of the microscopic density\ (\ref{dens}).
For a given volume $V$, $\widetilde{w}(k)$ vanishes for any $k\neq 0$ of the Fourier space $\Lambda$ in the limit $\alpha \to 0$ while $\widetilde{w}(0)=\widetilde{\Phi}(0)$. Therefore the potential  energy $E_p(\mathcal{C})$ of any GC configuration $\mathcal{C}$ will be
\begin{eqnarray}
-E_p(\mathcal{C})/k_B T&=& \frac{1}{2}\frac{1}{V}\widetilde{\rho}(0)^2  \widetilde{\Phi}(0) \nonumber \\
 &=&\frac{V}{2}\beta \rho^2 \; ,
\end{eqnarray}
where we have changed our notations and defined the reduce inverse temperature as $\beta\equiv \widetilde{\Phi}(0)$ and $\rho=N/V$ is the number density in configuration  $\mathcal{C}$. The GCPF of the model is then given (for a homogeneous system) by
\begin{equation}
\Xi_{V}^{\mathrm{Kac}}\left[ \nu \right]\simeq \int_{0}^{\infty} d\rho \;
\exp\left( -V \left( - \nu \rho + f_{\mathrm{HS}}\left( \rho\right) -\frac{1}{2} \beta \rho^2\right) \right) \; ,
\end{equation}
where we have replaced the sum over $N$ in equation\ (\ref{csi}) by an integral over the density which is valid for large systems with no consequences on the thermodynamic limit and   $f_{\mathrm{HS}}(\rho)$ is the specific Helmoltz free energy  of the HS fluid at density $\rho$. Note also that, in the limit $\alpha \to 0$, one has $\overline{\nu}=\nu$ since the self-energy $w(0)$ vanishes. Only now we take the limit $V \to \infty$ with the result

\begin{equation}
\label{poi}
- \lim_{V \to \infty} V^{-1}\ln \Xi_{V}^{\mathrm{Kac}}\left[ \nu \right]=  \min_{\rho}\left\lbrace
\mathcal{L} (\nu,\rho) \right\rbrace \; ,
\end{equation}
where
\begin{eqnarray}
\mathcal{L} (\nu,\rho)&\triangleq & f_{\mathrm{MF}}(\rho,\beta) - \rho \nu \; ,\nonumber \\
f_{\mathrm{MF}}(\rho,\beta)&\triangleq &f_{\mathrm{HS}}(\rho) - \frac{\beta}{2} \; \rho^2 \; .
\end{eqnarray}
The function $\mathcal{L} (\nu,\rho)$ of equation\ (\ref{poi})  plays therefore the role of a Landau function \cite{Goldenfeld}. Note that $f_{\mathrm{MF}}(\rho,\beta)$ is precisely the MF free energy as defined at equation\ (\ref{ineq}). 
The HS free energy $f_{\mathrm{HS}}(\rho)$ which enters the expression of $\mathcal{L} (\nu,\rho)$ is known analytically only in dimensions $D=1$ and $D=\infty$, however in the case $D=3$ -to which we stick to from now- good approximations are known, for instance the Carnahan-Starling (CS) approximation \cite{Hansen} for which we have  
\begin{equation}
\label{CS}
f_{\mathrm{HS}}(\rho)=\rho\left( \ln \rho \lambda^{3} -1\right) + \frac{6\eta^2\left( 4 -3 \eta \right) }{\pi \left( 1 -\eta\right)^2 } \; ,
\end{equation}
where $\lambda$ is de Broglie wavelength and $\eta=\pi \rho\sigma^3/6$ is the packing fraction. The CS approximation describes a single fluid phase from $\eta=0$ up to the unphysical $\eta=1$ and the function of equation\ (\ref{CS}) is strictly convex, i.e. its second derivative  $f_{\mathrm{HS}}^{(2)}(\rho)>0$ for all $0<\eta<1$. In fact the function $\rho \mapsto f_{\mathrm{HS}}^{(2)}(\rho)>0$ is itself convex with a unique positive minimum at $\rho_0 \sigma^3=0.2491295\ldots $  with  $\beta_c\triangleq f_{\mathrm{HS}}^{(2)}(\rho_0)=11.101579\ldots \;$ \cite{Cai-Mol}.
It follows from these remarks that for $\beta < \beta_c$ the Landau function 
$\rho \mapsto \mathcal{L} (\nu,\rho)$ is convex and coercive for any $\nu$ (since $\partial^2 f_{\mathrm{MF}}(\rho,\beta)/\partial \rho^2>(\beta_c -\beta) $ for all $\rho$) and that the minimum of $\mathcal{L} (\nu,\rho)$ in equation\ (\ref{poi}) exists and is unique. Conversely for 
$\beta > \beta_c$, $\mathcal{L} (\nu,\rho)$ admits in general two minima and there is only but one value of the chemical potential $\nu= \nu_{\mathrm{coex}}(\beta)$ for which the two minima $\rho_g(\beta)$ and $\rho_l(\beta)$  ($\rho_l > \rho_g$) which describes the gas and the liquid phases respectively are such that $\mathcal{L} (\nu_{\mathrm{coex}},\rho_g) = \mathcal{L} (\nu_{\mathrm{coex}},\rho_l)=-\beta P_{\mathrm{coex}}(\beta)$ where $\beta P_{\mathrm{coex}}$ is the pressure at coexistence. At $\beta_c$ the two densities $\rho_g$ and $\rho_l$ merge to $\rho_0$ and thus $\beta_c$ is the inverse critical temperature of Kac model and $\rho_0$ its  critical density. Note that for $\beta > \beta_c$ the Kac pressure has a kink at $\nu=\nu_{\mathrm{coex}}$ with two distinct slopes $\rho_g$ and $\rho_l$. To this kink corresponds a segment for the Legendre transform, i.e. the GC free energy\cite{Goldenfeld}. One thus have, for $\beta<\beta_c$
\begin{eqnarray}
\label{jojo1}
f_{\mathrm{Kac}}(\rho,\beta) &=& f_{\mathrm{MF}}(\rho,\beta)  \;,
\end{eqnarray}
and for $\beta > \beta_c$
\begin{eqnarray}
\label{jojo2}f_{\mathrm{Kac}}(\rho,\beta) &=& f_{\mathrm{MF}}(\rho,\beta) \; \; \mathrm{ for } \; \;
\rho > \rho_l(\beta) \; \; \mathrm{ or } \; \; \rho < \rho_g(\beta) \nonumber \; , \\
 &=& f_{\mathrm{MF}}(\rho_g,\beta)\frac{\rho -\rho_l}{\rho_g -\rho_l} +
f_{\mathrm{MF}}(\rho_l,\beta)\frac{\rho -\rho_g}{\rho_l -\rho_g} \; \; \mathrm{ for } \; \; \rho_g < \rho < \rho_l \; ,
\end{eqnarray}
where $\rho_g(\beta)$, $\rho_l(\beta)$, and $\nu_{\mathrm{coex}}(\beta)$ are the unique solution of the system
\begin{eqnarray}
\label{system}f_{\mathrm{MF}}(\rho_g,\beta) -\nu_{\mathrm{coex}}\rho_g &=&f_{\mathrm{MF}}(\rho_l,\beta) -\nu_{\mathrm{coex}}\rho_l \; ,\nonumber \\
\partial f_{\mathrm{MF}}(\rho_g,\beta)/\partial \rho_g&=&\partial f_{\mathrm{MF}}(\rho_l,\beta)/\partial \rho_l=\nu_{\mathrm{coex}} \; .
\end{eqnarray}
\subsection{\label{nprg-Kac} NPRG for Kac model}
\begin{figure}
\includegraphics[angle=0,scale=0.6]{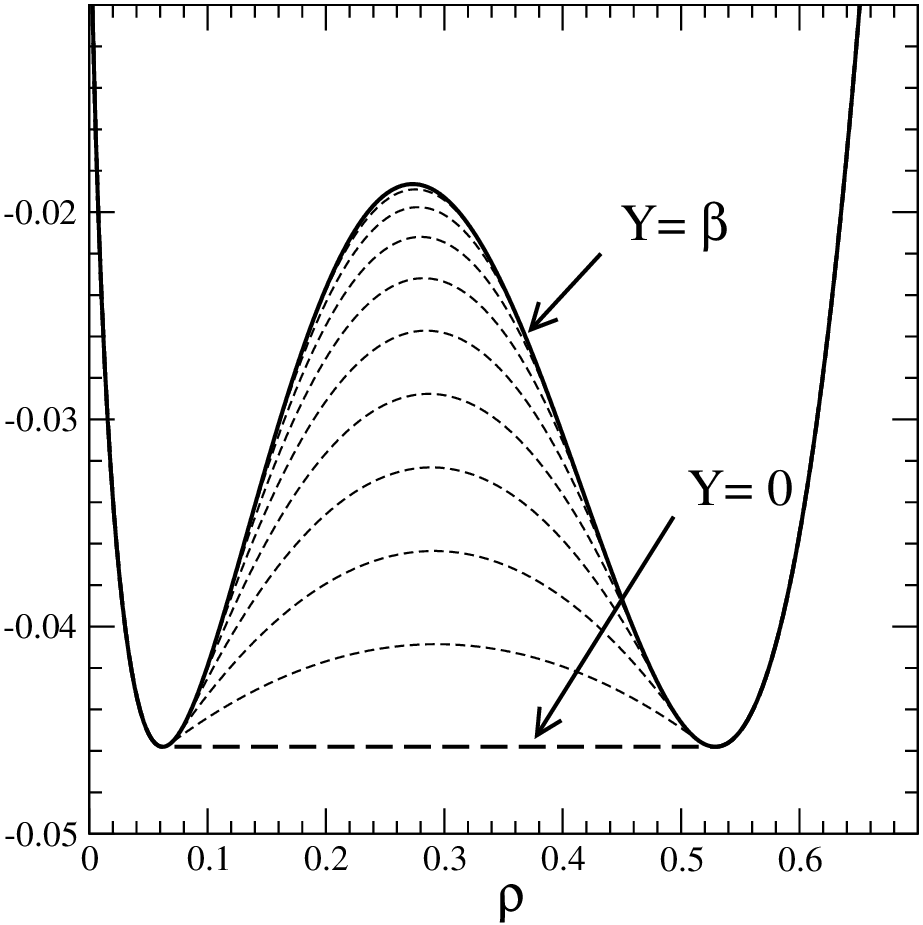}
\caption{\label{fig1} Flow of  $f_Y(\rho)-\nu_{\mathrm{coex}}(\beta) \rho$ at $\beta=13 > \beta_c$. At $Y=\beta$ (thick solid line) $f_Y\equiv f_{\mathrm{MF}}$, at $Y=0$ (thick dashed line) $f_Y\equiv f_{\mathrm{Kac}}$. As $Y$ decreases (dashed lines from top to bottom) $f_Y$ becomes convex.}
\end{figure}

Our starting point for the NPRG analysis of Kac model will be equation\ (\ref{flow-f}) that we rewrite as
\begin{equation}
\label{flow-fk2}
\partial f_k(\rho) /\partial k^2 =
 \frac{1}{2} \int_{q} \frac{\partial \widetilde{w}_{k}(q)/ \partial k^2
}{\widetilde{C}_{k}^{(2)} (q) + \widetilde{w}_{k}(q) - \widetilde{w}(q)} \; .
\end{equation}
We do not aim at mathematical rigor and will solve the flow by hand-waving arguments. Taking the Kac limit $\alpha \to 0$ at fixed $V$  in the RG framework means that at any scale $k\neq 0$ one has $k\gg\alpha$ and $q \gg \alpha$, therefore
\begin{equation}
\label{zo1}\widetilde{w}(q)\simeq \left( 1 - \Theta_{\alpha}\left( q\right) \right) \beta \; , 
\end{equation}
where $\Theta_{\alpha}\left( q\right)$ is a smoothened step function equal to $1$ for $q>\alpha$ and to $0$ for $q<\alpha$. Whatever the form of the cut-off function $R_k(q)$ one thus has 
\begin{eqnarray}
\label{zo2}\widetilde{w}_k(q)&\simeq &\left( 1 - \Theta_{\alpha}\left( q\right) \right) \frac{\beta}{1+ k^2 \beta} \; , \nonumber \\
\partial_{k^2} \widetilde{w}_{k}(q)&\simeq & \left( 1 - \Theta_{\alpha}\left( q\right) \right) \frac{-\beta^2}{\left( 1+ k^2 \beta\right)^2 } \; ,
\end{eqnarray}
since $R_k(q)\sim k^2$ for $q\to 0$ (cf the conditions\ (\ref{111}). Because of the presence of the $\left( 1 - \Theta_{\alpha}\left( q\right) \right)$ in the numerator of the right hand side of equation\ (\ref{flow-fk2}) one can replace the function $\widetilde{C}_{k}^{(2)} (q)$ at the denominator by its value at $q=0$, i.e. $-\partial^{2} f_k(\rho)/\partial \rho^2$. Therefore for $\alpha \ll k$ one has the simplified flow equation
\begin{equation}
\partial f_k(\rho) /\partial k^2 = \frac{4 \pi}{6} \alpha^3 \frac{\beta^2}{\left(1 + k^2 \beta \right)\left(\partial^2f_{k}/\partial \rho^2 + \frac{k^2 \beta^2}{1 + k^2 \beta^2}  \right)  } \; ,
\end{equation}
which can be rewritten, with the change of variable
\begin{equation}
Y=\frac{k^2 \beta^2}{1+ k^2 \beta} \; ,
\end{equation}
as
\begin{equation}
\partial f_Y(\rho)/\partial Y \left[ \partial^2 f_Y(\rho) /\partial \rho^2 + Y \right]= \frac{4 \pi}{6} \alpha^3 \; ,
\end{equation}
where $f_Y(\rho) \equiv f_k(\rho)$. One can  take safely the limit $\alpha \to 0$ of the equation above which yields
\begin{equation}
\label{flowY}\partial f_Y(\rho)/\partial Y \left[ \partial^2 f_Y(\rho) /\partial \rho^2 + Y \right]= 0 \; .
\end{equation}
This somehow strange partial differential equation  must be solved with the initial conditions\ (\ref{cond-init}), i.e at $k=\infty \Leftrightarrow Y=\beta$ one has $f_{Y=\beta}(\rho)=f_{\mathrm{MF}}(\rho,\beta)=f_{\mathrm{HS}}(\rho)-\beta \rho^2/2$. 
\begin{figure}
\includegraphics[angle=0,scale=0.6]{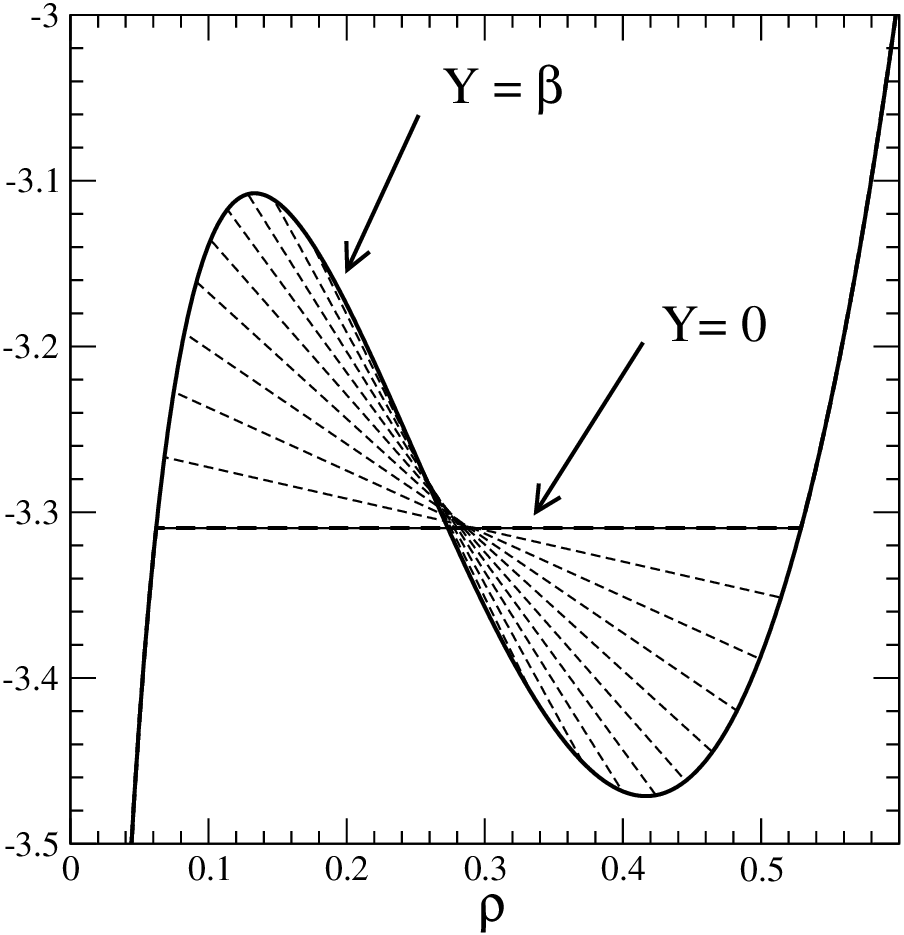}
\caption{\label{fig2} Flow of  $\partial f_Y(\rho)/\partial \rho$ at $\beta=13 > \beta_c$. As $Y$ decreases (dashed lines from top to bottom) a plateau appears.}
\end{figure} 
Let us define $X\triangleq Y +  \partial^2 f_Y(\rho) /\partial \rho^2$. We first remark that, at the beginning of the flow, i.e. for $Y=\beta$,  the initial conditions are such that that $X=f^{(2)}_{\mathrm{HS}}(\rho) > \beta_c \equiv f^{(2)}_{\mathrm{HS}}(\rho_0)>0$ for all $\rho$. It follows then from the flow equation\ (\ref{flowY}) that $\partial f_Y(\rho)/\partial Y = 0 $ and therefore the flow does not start immediately. In this initial stage of immobility $X=f^{(2)}_{\mathrm{HS}}(\rho)-\beta +Y \geq ( \beta_c -\beta +Y)$ for all $\rho$, while $\partial f_Y(\rho)/\partial Y = 0 $ if $\beta_c -\beta +Y>0$.

Now, if $\beta < \beta_c$ then $X$ remains strictly positive for all $\rho$ when $Y$ varies from $Y=\beta$ ($k=\infty$) to $Y=0$ ($k=0$). Therefore $\partial f_Y(\rho)/\partial Y $ remains zero and the flow never starts. We thus have $f_{Y=0}(\rho)=f_{\mathrm{MF}}(\rho,\beta)=f_{\mathrm{Kac}}(\rho,\beta)$ as expected.

The case $\beta > \beta_c$ is more interesting. For $\overline{Y}=\beta -\beta_c$ the minimum of $X(\rho)$ becomes equal to zero and the flow can start. If it actually does then it follows from \ (\ref{flowY}) that $ Y +  \partial^2 f_Y(\rho) /\partial \rho^2=0$ for $0<Y<\overline{Y}$ in some domain of $\rho$. It seems reasonable to assume that, at  given $Y$, this domain is some interval $[\rho_1(Y), \rho_2(Y)]$. Therefore
\begin{equation}
f_Y(\rho)=-\frac{1}{2}Y \rho^2 + a(Y) \rho + b(Y) \; 
\end{equation}
for $\rho_1<\rho<\rho_2$. For $\rho >\rho_2$ or $\rho<\rho_1$ the flow has not yet started and $f_Y$ remains equal to its initial value $f_{\mathrm{MF}}(\rho,\beta)$. In order to determine the unknown functions $a(Y)$, $b(Y)$, $\rho_1(Y)$ and $\rho_2(Y)$ we invoke the continuity of the functions $\rho\mapsto f_Y(\rho)$ and $\rho\mapsto \partial f_Y/(\rho)/\partial \rho$ which can be inferred from the general flow equation\ ( \ref{flow-fk2}). Hence we have $\forall Y\in [0,\overline{Y}]$
\begin{eqnarray}
\label{lol}
f_{\mathrm{MF}}(\rho_i,\beta)&=& -\frac{1}{2}Y \rho_i^2 +a(Y)\rho_i +b(Y) \; \; (i=1,2) \; , \nonumber \\
\partial f_{\mathrm{MF}}(\rho_i,\beta)/\partial \rho_i &=& -Y \rho_i +a(Y)  \; \; (i=1,2) \; .
\end{eqnarray}
In order to solve equations\ (\ref{lol}) we remark that the continuity conditions\ (\ref{lol})  can  be reformulated as $f_{\mathrm{MF}}(\rho_i,\Delta)=a(Y) \rho_i +b(Y) $ and
$\partial f_{\mathrm{MF}}(\rho_i,\Delta)/\partial \rho=a(Y)$ 
with $\Delta=\beta-Y$.  These relations are precisely those which determine the coexistence curve of the Kac model at the inverse temperature $\Delta$ (cf equations\ (\ref{system})).
Therefore
\begin{eqnarray}
\rho_1(Y)&=&\rho_g(\Delta) \; , \nonumber \\
\rho_2(Y)&=&\rho_l(\Delta) \; .
\end{eqnarray}
In other words, the flows of $\rho_1(Y)$ and $\rho_2(Y)$ are given by the two branches of the binodal of the Kac model at inverse temperature $\Delta=\beta -Y$. The starting point of the flow $Y=\overline{Y}$ corresponds to $\Delta=\beta_c$, i.e to the inverse critical temperature of Kac model. Moreover  $a(Y)\equiv \nu_{\mathrm{coex}}(\Delta)$ coincides with the coexistence chemical potential of the Kac model at inverse temperature $\Delta$ and  $b(Y)=-\beta P_{\mathrm{coex}}(\Delta)$ with minus the  pressure at  coexistence. Our final result can be summarized as
\begin{eqnarray}
\label{solu-flow}
f_Y(\rho) &=&f_{\mathrm{MF}}(\rho,\beta ) \; \;  \mathrm{for} \; \;
\rho<\rho_g(\Delta) \; \; \mathrm{or}\; \;  \rho>\rho_l(\Delta) \; , \nonumber \\
&=& -\frac{1}{2} Y \rho^2 + \nu_{\mathrm{coex}}(\Delta) \rho - \beta P_{\mathrm{coex}}(\Delta)\; \;  \mathrm{for} \; \;
\rho_g(\Delta) < \rho<\rho_l(\Delta) \; .
\end{eqnarray}
At the end of the flow, i.e. for $Y=0$, one has $\Delta=\beta$ and thus 
$f_Y(\rho)\equiv f_{\mathrm{Kac}}(\rho,\beta)$ as given by equations\ (\ref{jojo1}) and\ (\ref{jojo2}). 

To conclude this section let us have a look on the flow of the "true" free energy $f_k^{'}(\rho)$. It follows from equation\ (\ref{def-Ak}) that
\begin{eqnarray}
f_k^{'}(\rho)&=& f_k(\rho) + \frac{1}{2}\rho^2 \left(\widetilde{w}(0) -\widetilde{w}_k(0) \right) \; , \nonumber \\
&=&f_k(\rho) +\frac{1}{2} Y \rho^2 \; ,
\end{eqnarray}
from which we deduce that $f_Y^{'}(\rho)\equiv f_k^{'}(\rho)=f_{\mathrm{Kac}}(\rho,\Delta)$ which is an obvious result since the k-system is nothing but a Kac model at inverse temperature $\Delta$
as can be inferred from equations\ (\ref{zo1}) and\ (\ref{zo2}). This could have been our starting point to solve the flow. This allows to find the exact solution of the flow without solving explicitely equation\ (\ref{flow-fk2}), but it is less funny.

To illustrate our result we have displayed the flow of the running free energy
and its two first derivatives with respect to $\rho$. Figure 1 shows the flow of $f_Y(\rho)-\nu_{\mathrm{coex}}(\beta) \rho$ at a temperature $T^*=1/\beta$ below $T_c$ as $Y$ varies from $Y=\beta$ to $Y=0$. This demonstrates the approach of convexity in the inner region when $Y \to 0$. Figure 2 displays the flow of $\partial f_Y(\rho)/\partial \rho $ at the same subcritical temperature, demonstrating the slow construction of the plateau. Finally, in Figure 3 we displays the flow of $\partial^2 f_Y(\rho)/\partial \rho^2$ which exhibits a finite discontinuity at the points $\rho_1(Y)=\rho_g(\Delta)$ and  $\rho_2(Y)=\rho_l(\Delta)$. Since at $Y=0$, the isothermal compressibility $\chi_{T}$ given by $\beta \chi_{T}^{-1}=\rho^{-2}\partial^2 f_{Y=0}(\rho)/\partial \rho^2$, $\chi_{T}$ is finite on the binodal although infinite in the coexistence region. 

We end this section by pointing out that the HRT flow cannot be solved conveniently for the Kac model, since at any scale $k\neq 0$ we have $\widetilde{w}_k(q)\equiv 0 \; \; \forall q $, as a consequence of the ultra sharp cut-off. In this case the flow starts and stops exactly at $k=0$ with an uncontrollable singularity.

\begin{figure}
\includegraphics[angle=0,scale=0.6]{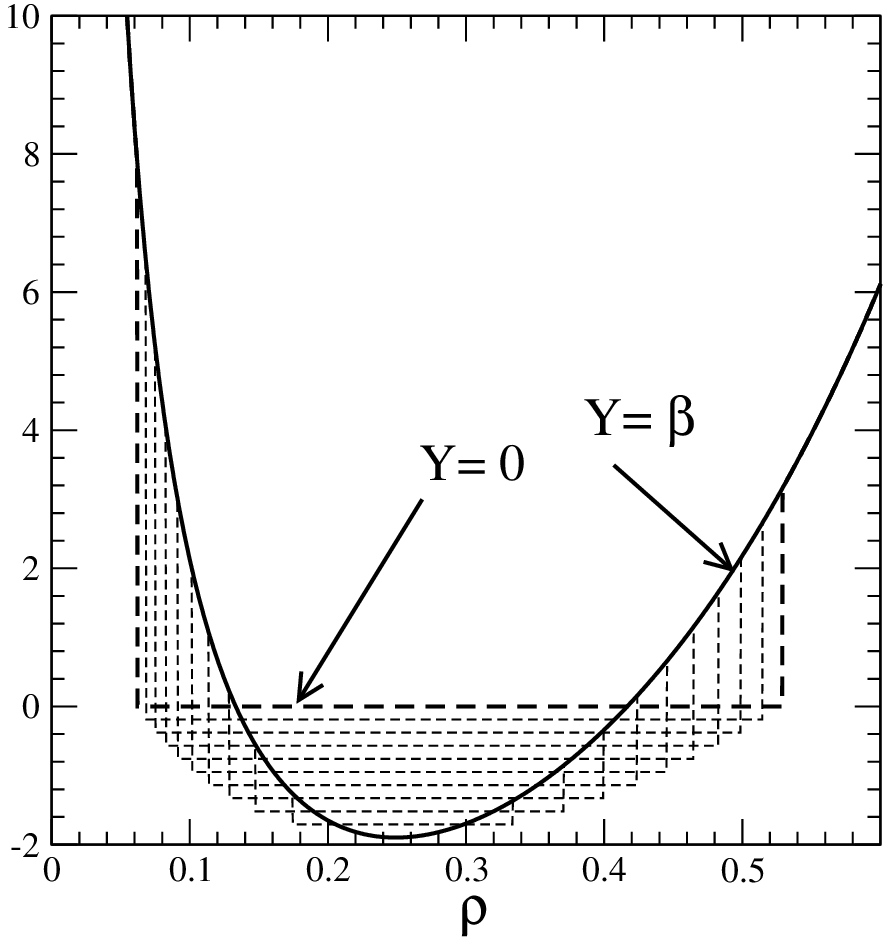}
\caption{\label{fig3} Flow of  $\partial^2 f_Y(\rho)/\partial \rho^2$ at $\beta=13 > \beta_c$. The compressibility is discontinuous as soon as the flow starts.}
\end{figure} 
\section{\label{conclusion}Conclusion}
In this paper we have developed a smooth cut-off NPRG theory for simple fluids. Two equivalent versions of the theory were formulated, the first one in the GC ensemble, with an immediate transcription in the framework of KSSHE field theory, the second one in the framework of a generic scalar field theory (with no Z2 symmetry). The exact mapping between these two versions was derived in detail. The HRT is obtained as a highly singular limit of regular smooth cut-off NPRG equations. The inability of HRT to distinguish between the spinodal and binodal curves is probably a consequence of this limit which is only partly controllable (the flow equations for the $C^{(n)}_k$ with $n\geq 2$ are ill-defined in the scaling limit $k\to 0$). This shortcoming is absent of the present theory.

Our developments where highly abstract and formal and we did not say anything on the strategy to adopt to solve the NPRG flow equations so some comments on this issue are probably wellcome. There are mainly two ways to solve the NPRG equations. A first method is to restrict the functional form of the  effective average action $\Gamma_k[\Phi]$ (or $\beta \mathcal{A}_k[\rho]$) on a restricted functional sub-space. For instance the simplest  ansatz is the so-called local potential (LPA) approximation\cite{Wetterich,Bervillier,Delamotte}
\begin{equation}
\label{LPA}\Gamma_k[\Phi]=\int_x U_k\left( \left(\Phi(x) \right) \right) +\frac{1}{2}
\left[ \nabla \Phi(x)\right]^2 
\end{equation}
where $U_k\left( \left(\Phi(x) \right) \right) $ is a \textit{function} of the classical field. The LPA ansatz allready gives non trivial critical exponents (except $\eta=0$) and is quite easy to solve numerically\cite{Wetterich}. Allready the improvement $1/2 \rightarrow Z_k/2$ in equation\ (\ref{LPA}) which includes the renormalization of the wave function and yields $\eta\neq 0$ makes the results more satisfactory.
It must be precised first that most of the numerical studies of LPA or improved LPA equations were applied to systems with Z2 symmetry, i.e. with no direct application to liquids, see however reference\cite{Litim} where symmetric and antisymmetric correction-to-scaling exponents are studied in detail. 
Secondly, these ansatz (including those with more terms in the gradient expansion) allow to estimate the low-q behavior of \textit{all} the vertex but do not give their correct behavior at short distances in direct space. Such methods will not fit with the dogma of the theory of liquids according to which $g(r)$ should be zero in the core. In liquids the equation of state (i.e. the virial pressure) results from a delicate balance between energetic and entropic contributions (i.e. the value of the $g(r)$ at contact, that is at $r=\sigma$). If applied to liquid theory, approximations such as LPA or gradient expansions although giving good estimations of critical exponents will miss this point and would yield a poor estimation of non universal critical properties. This is why Reatto and coworkers considered another strategy to solve their HRT hierarchy. It consists in closing the tower of equations for the $C_k^{(n)}$. Contrary to the first strategy high order $C_k^{(n)}$ are ignored but the structure of $C_k^{(2)}(r)$ can be described with good accuracy for all "$r$". In particular one can take advantage of the arbitrariness of $w_k(r)$ in the core to enforce $g_k(r)\equiv 0$ for $0<r<\sigma$ at each scale-k. The most recent HRT studies\cite{Reiner1,Reiner2,Reiner3,Reiner4} consider a closure in the spirit of the Optimized RPA (ORPA)\cite{Chandler}
\begin{eqnarray}
C_k^{(2)}(r)&=& C_{\mathrm{HS}}^{(2)}(r)+\gamma^0_k w(r) + K_k(r) \nonumber \; , \\
K_k(r)&=&\sum_{n=1}^{n_{max}}\gamma^n_k u^n(r) \; , 
\end{eqnarray}
where the $u^n(r)$ are taken to provide a basis for a suitable function space over $[0,\sigma]$. Coefficient $\gamma^0_k$ is chosen to enforce thermodynamic consistency (i.e $\widetilde{C}_k^{(2)}(0)=-\partial^2 f_k /\partial \rho^2$) while the $n_{max}$ and the $\gamma^n_k$ will be determined in such a way that $g(r)$ is zero in the core with sufficient accuracy. It seems that this numerical scheme could be adapted to our NPRG flow equations, but other schemes proposed recently for solving Wetterich flow should also be considered\cite{Wschebor,Wschebor2}. We hope to report progresses on these points in future works.

\begin{acknowledgments}
I attended the lectures of  B. Delamotte on the NPRG in spring 2005 at Jussieu University (Paris-France) which was a marvelous introduction to the subject. The interested reader is urged to consult his lecture notes  which are  available on the Web \cite{Delamotte}.
\end{acknowledgments}

\end{document}